\documentclass[a4paper,11pt]{article}
\pdfoutput=1 % if your are submitting a pdflatex (i.e. if you have
\usepackage{jcappub} % for details on the use of the package, please
\usepackage{enumerate}
\usepackage{tikz}
\usepackage{enumitem}
\usepackage{booktabs}
\usepackage{xspace}
\usepackage{subcaption} % for subfigure
\graphicspath{figs/}
\usepackage[justification=raggedright,singlelinecheck=false]{caption}
\usetikzlibrary{arrows,positioning}

\usepackage{aas_macros} % For journal abbreviations

\newcommand{\gr}{$\gamma$-ray\xspace}
\newcommand{\grs}{$\gamma$-rays\xspace}
\newcommand{\sgr}{Sgr $\mathrm{A^*}$}
\newcommand{\Fermi}{\textit{Fermi}}

\begin{document}

\title{Prospects for indirect dark matter searches with MeV photons}

\author[a]{Richard Bartels,}
\author[a]{Daniele Gaggero}
\author[a]{and Christoph Weniger}

\affiliation[a]{Gravitation Astroparticle Physics Amsterdam (GRAPPA),
Institute for Theoretical Physics Amsterdam 
and Delta Institute for Theoretical Physics,\\
University of Amsterdam, Science Park 904, 1098 XH Amsterdam, 
The Netherlands}

% e-mail addresses: one for each author, in the same order as the authors
\emailAdd{r.t.bartels@uva.nl}
\emailAdd{d.gaggero@uva.nl}
\emailAdd{c.weniger@uva.nl}

\date{\today}

\abstract{
  Over the past decade, extensive studies have been undertaken to search for
  photon signals from dark matter annihilation or decay for dark matter
  particle masses above $\sim1$ GeV.  However, due to the lacking sensitivity
  of current experiments at MeV--GeV energies, sometimes dubbed the `MeV gap',
  dark matter models with MeV to sub-GeV particle masses have received little
  attention so far.  Various proposed MeV missions
  (like, \textit{e.g.}, e-ASTROGAM or AMEGO) are aimed
  at closing this gap in the mid- or long-term future.  This, and the 
  absence of clear dark matter signals in the GeV--TeV range, makes it relevant
  to carefully reconsider the expected experimental instrumental sensitivities
  in this mass range.  The most common two-body annihilation channels for
  sub-GeV dark matter are to neutrinos, electrons, pions or directly to
  photons.  Among these, only the electron channel has been extensively
  studied, and almost exclusively in the context of the 511 keV line.  In this
  work, we study the prospects for detecting MeV dark matter annihilation in
  general in future MeV missions, using e-ASTROGAM as reference, and focusing
  on dark matter masses in the range 1\,MeV--3\,GeV.  In the case of leptonic
  annihilation, we emphasise the importance of the often overlooked
  bremsstrahlung and in-flight annihilation spectral features, which in many
  cases provide the dominant gamma-ray signal in this regime.
}


\maketitle

%%%%%%%%%%%%%%%%%%%%%%%%%%%%%%%%%%%%%%%%%%%%
%%%%%%%%%%%%%%%%%%%%%%%%%%%%%%%%%%%%%%%%%%%%

%-----------------------------------------------------------------------------------------
% INTRODUCTION
%-----------------------------------------------------------------------------------------

\section{Introduction}

The past two decades have seen a rapid development of \gr astronomy. The
Spectrometer on Integral (SPI) has clearly detected and characterised the
diffuse 511 keV electron-positron annihilation signal \citep{Winkler:2003nn,
Vedrenne:2003, Knodlseder:2005yq, Weidenspointner:2006nua, Siegert:2015knp},
and the \Fermi\ Large Area Telescope (LAT) has revolutionized the field at
energies above $\mathcal{O}(1\mathrm{\,GeV})$ \cite{Atwood:2009ez}.  Yet,
sensitivity in the MeV--GeV range has notoriously trailed behind.  Existing
measurements of diffuse $\gamma$-rays were taken with COMPTEL and EGRET
\cite{Strong:1998ck, Hunter:1997we}. A number of instruments have been
proposed over the last years to cover this so-called `MeV gap'
\cite{Knodlseder:2016pey}. Currently, the community focuses on various
proposals, like e-ASTROGAM\footnote{e-ASTROGAM is proposed as ESA M5 mission.}
\cite{DeAngelis:2016slk} and AMEGO\footnote{See
\url{https://pcos.gsfc.nasa.gov/physpag/probe/AMEGO_probe.pdf}.} (other
previously proposed missions include, \textit{e.g.}, COMPAIR
\cite{Moiseev:2015lva} and ADEPT \cite{Hunter:2013wla}), which could be
realized in the mid- and long-term future in the late 2020s.  Whatever
instrument is ultimately realized, it is expected that it would improve the
sensitivity of current dark matter searches with MeV \grs by 2--3 orders of
magnitude~\cite{Essig:2013goa, Boddy:2015efa}.

Annihilating or decaying dark matter (DM) can leave observable features in the
electromagnetic spectrum (for a review see Ref.~\cite{Bringmann:2012ez}, and
references therein).  One goal of the proposed MeV telescopes would be to look
for signals from particle DM in the diffuse MeV sky. In particular, these
telescopes would be sensitive to DM with masses in the MeV--GeV range,
henceforth `MeV DM'.  
At this mass scale, the number of kinematically allowed
final states becomes limited to neutrinos, pions, photons and light leptons.
For DM with masses $\lesssim 10\mathrm{\,MeV}$ in thermal equilibrium in the
early universe the strongest constraints come from big-bang nucleosynthesis
(BBN), where it can alter the elementary abundances, and from the
cosmic-microwave background (CMB) where it affects the effective number of
neutrino species. \citep[e.g.,][]{Serpico:2004nm, Boehm:2013jpa,
Nollett:2013pwa, Nollett:2014lwa}.  These constraints are independent of
whether DM annihilates via $s$- or $p$-wave.  In addition, late-time energy
injection from DM, between the epochs of recombination and reionization, can
leave a strong imprint in the CMB \cite[e.g.,][]{Galli:2009zc, Ade:2015xua,
Slatyer:2015jla}. However, since thermal velocities at recombination are low,
such bounds are only strong for $s$-wave annihilating DM.  Models of DM with
MeV masses that \textit{can} satisfy these constraints and yield the correct
relic density include self-interacting DM \cite{Boehm:2003hm, Pospelov:2007mp,
DAgnolo:2015ujb, Chu:2016pew, Choi:2016tkj}, `cannibal'
DM~\cite{Pappadopulo:2016pkp} and strongly-interacting DM
\cite{Hochberg:2014dra}.

Detecting MeV DM is challenging in a number of ways. Direct detection
experiments based on nuclear recoils are not sensitive below a GeV, and bounds
from ionization are typically weak \cite{Essig:2012yx}, although some
alternative detection principles have the potential to substantially improve
current sensitivities in the future \citep[e.g.,][]{Essig:2013goa,
Essig:2015cda, Hochberg:2015pha}. Electrons and positrons from MeV DM annihilation or decay are very hard to detect through measurements of the top-of-atmosphere positron fraction 
because solar modulation cuts-off the
cosmic-ray (CR) spectrum at $\mathcal{O}(1 \mathrm{\,GeV/nucleon})$
\cite{longair2011high}.  A notable exception is here Voyager 1, which has left
the Heliosphere, and can constrain MeV DM through measurements of the
local lepton spectrum~\cite{Boudaud:2016mos}.  

However, indirect detection of the gamma-ray emission from (leptonic) DM
annihilation or decay is possible \citep[e.g.~][]{Boehm:2002yz}. 
One notable example is the 511 keV anomaly,
which has been explained in terms of DM annihilating to $e^+e^-$-pairs
(see e.g.~\cite{Boehm:2003bt} or, for a review, \cite{Prantzos:2010wi} and
references therein).
However, many models
considered in this context (based on light, thermally produced WIMPs) are now
ruled out by the early-universe bounds mentioned above~
\cite{Wilkinson:2016gsy}. Thus, a DM explanation of the 511 keV line may
require for instance non-thermal or exciting DM \cite{Wilkinson:2016gsy,
Finkbeiner:2007kk}.

One promising general way to look for MeV DM is to search for spectral features
in diffuse $\gamma$-rays. These features can come from the prompt
$\gamma$-rays produced in the annihilation, or they are related to secondary
emission in leptonic annihilation channels (additional spectral features at
sub-GeV energies can come from the decay of meson excited states, as pointed
out recently in Ref.~\cite{Bringmann:2016axu}).  Conservative upper limits on
prompt radiation were presented in Refs.~\cite{Essig:2013goa, Boddy:2015efa},
without subtraction of any diffuse backgrounds from the data.  Moreover,
leptonic final states like $e^+e^-$-pairs can play a particularly interesting
role.  Aside from the prompt final-state radiation (FSR) signature, there will
be a large flux of secondary $\gamma$-rays, dominated by in-flight annihilation
(IfA) of positrons.  The relevance of IfA was pointed out long ago in
Refs.~\citep*{Aharonian:1981spy, Aharonian_1983, Aharonian:2000iz}, and it has
been used to constrain DM explanations of the 511 keV line signal
\cite{Beacom:2005qv, Sizun:2006uh}.  In addition, there will be a sizeable
bremsstrahlung signal. Bremsstrahlung is often ignored in computations of the
\gr spectrum from WIMPS, but it becomes increasingly important when going to
lower DM masses \cite{Cirelli2013mqa}.

\medskip

In this paper, we study the prospects for a future $\gamma$--ray experiment to
detect DM through diffuse $\gamma$--rays originating in the Galactic halo.  We
will use the characteristics of the proposed e-ASTROGAM
\cite{DeAngelis:2016slk} as reference.  However, we emphasize that our
results are also representative for other similar missions, and comment on
how the results change with observation time, effective area or energy
resolution where appropriate. For the first time, we
perform an elaborate study of the detection opportunities of the secondary
emission in case of DM annihilation into leptons.  In addition, we attempt to
derive more optimistic, but also more realistic, projected upper limits for the
phenomenologically most interesting final states by modeling the expected
uncertainties in the diffuse background.  To this end, we assume that remaining
systematic uncertainties in these future missions will be of similar size as
the ones from the \Fermi-LAT today.  
These uncertainties are then incorporated into a novel statistical approach based on
Fisher forecasting~\cite{Edwards:2017mnf} (see also Ref.~\cite{Bringmann:2016axu})
which we apply to derive projected upper limits.
We concentrate here on the a region around the Galactic center, since we know
from the \Fermi-LAT and H.E.S.S.~\cite{Bringmann:2012vr, Ackermann:2013uma,
Ackermann:2015lka, Abdallah:2016ygi} that this region provides the best (while
still reasonably robust) probe for spectral signatures from DM annihilation.

The structure of this paper is as follows: in sections \ref{sec:models} and
\ref{sec:secondary} we discuss the different annihilation channels and their
resulting \gr spectra. Sec.~\ref{sec:secondary} is fully devoted to the
secondary \grs.  We discuss background modeling and the Fisher formalism used
to calculate upper limits in Sec.~\ref{sec:sensitivity}. 
Projected upper
limits are presented in Sec.~\ref{sec:results} and we end with a discussion and
conclusions in Sec.~\ref{sec:discussion} and \ref{sec:conclusion}.

%-----------------------------------------------------------------------------------------
% DM Models and Gamma-Ray fluxes
%-----------------------------------------------------------------------------------------
\section{Dark matter annihilation channels and photon signals}
\label{sec:models}

For MeV DM particles, $\chi$, only a few kinematically-allowed two-body
annihilation channels exist~\cite{Boddy:2015efa}.  We consider here the
following processes.
\begin{itemize}
  \item $\chi\chi\to\gamma\gamma$: A photon pair
  \item $\chi\chi\to\gamma\pi^0$: A neutral pion and a photon
  \item $\chi\chi\to\pi^0\pi^0$: Neutral pions
  \item $\chi\chi\to \bar\ell\ell$: Light leptons (with $\ell=e, \mu$)
  \item $\chi\chi\to \phi\phi$ and $\phi\to e^+e^-$: Cascade annihilation
\end{itemize}

We will here present a complete discussion of the phenomenology of all of these
channels in the MeV--GeV regime.  For instance, we provide updated prospects
for the detectability of DM models in which annihilation proceeds through
first-generation quarks into neutral pions or directly into gamma-ray lines as
discussed in \cite{Boddy:2015efa}.  However, much of the paper is focused on DM
annihilating into charged leptons, since the evaluation the expected signal is
more complex than in the other cases.  For each channel we will assume for
simplicity that the branching ratio is 100\%.

\subsection{General calculation of gamma-ray signal}

The $\gamma$-ray flux resulting from DM annihilation can be split up into two
components: a primary and a secondary component.

The \textit{primary} component is composed of all photons that are produced
directly in the annihilation process.  Their differential flux is given by
\cite[e.g.,][]{Cirelli:2010xx}
\begin{equation}
  \label{eq:flux}
  \frac{d\Phi}{dE d\Omega}=
  	\frac{a \left<\sigma v\right> J}{4\pi m_\chi^2}
    \frac{d N_\gamma}{dE_\gamma}\;,
\end{equation}
where $\left<\sigma v\right>$ is the velocity-averaged annihilation 
cross section, $J$ is the
astrophysical factor which encapsulates all information about the DM
distribution, $m_\chi$ is the DM mass, $dN/dE_\gamma$ the differential
$\gamma$-ray yield per annihilation and $a=\frac 1 2\, (\frac 1 4)$ if DM is
(is not) self-conjugate.  Throughout this work we assume that DM is
self-conjugate.

The so-called astrophysical-, or J-factor, is given by
\begin{equation}
  J=\int_\mathrm{l.o.s.} ds\, \rho^2(r(s,\theta))\;,
\end{equation}
where $\rho(r)$ is the DM density as function of the Galacto-centric radius
$r$. We adopt here a Navarro-Frenk-White (NFW) density profile
\cite{Navarro:1996gj}, $\rho_\mathrm{NFW}(r)=\rho_0/((r/r_s)^\gamma
(1+r/r_s)^{(3-\gamma)})$, with a local DM density of
$\rho_\odot=0.4\mathrm{\,GeV\,cm^{-3}}$ \citep[e.g.~][]{Read:2014qva, 
Catena:2009mf, Weber:2009pt, Salucci:2010qr, Iocco:2011jz, 
McMillan:2011wd}, scale radius $r_s=20\mathrm{\,kpc}$
and slope $\gamma=1$. Since DM annihilation scales as $\rho^2$, our results are
-- as typical for searches in the inner Galaxy -- very sensitive to the adopted
profile.  We will address how different assumptions on the profile affect our
main results in Sec.~\ref{sec:discussion}.

Regarding the DM annihilation cross section, given the usual decomposition
$\left(\sigma v\right)=a + b v^2$, with $a$ the $s$-wave and $bv^2$ the
$p$-wave contribution, we require that the $p$-wave term sets the relic density
and still dominates at the time of recombination, in order to avoid the CMB
constraint from late-time energy injection mentioned in the introduction.  For
this reason, we expect a value much below the canonical $3\cdot
10^{-26}\mathrm{\,cm^3\,s^{-1}}$.  For example,  for a thermal relic
  purely annihilating through $p$-wave processes, the expected annihilation
cross section times velocity in the Galaxy today is of the order 
$\sim10^{-31}\mathrm{\,cm^3\,s^{-1}}$ 
(see discussion in Sec.~\ref{sec:constraints}). 

\medskip

The prediction of the \textit{secondary} $\gamma$-ray flux from charged
particles (here electrons) is significantly more complex.  The $\gamma$-ray
spectrum will depend on the energy-losses of the electrons, and on the
radiative process underlying the emission. In addition, the morphology will no
longer trace the DM squared distribution directly, but rather depend on how far
and in what direction the electrons have propagated.  These environmental
impacts are discussed extensively in section~\ref{sec:secondary}.

%----------------------------
% Neutral pions & gamma ray lines
%----------------------------

\subsection{Gamma-ray lines and neutral pions}

We briefly summarize the analytical expressions for the various prompt
annihilation spectra that we consider in this work.

\begin{enumerate}[label=(\roman*)]
  \item $\chi\chi\rightarrow\gamma\gamma$.
    For this channel, the photon spectrum per annihilation is given by
    \begin{equation}
      \frac{dN_\gamma}{dE} = 2\delta(E-m_\chi)\;.
    \end{equation}

  \item $\chi\chi\rightarrow\pi^0\pi^0$.
    The neutral-pion channel leads to a box-shaped $\gamma$-ray spectrum
    \cite{Ibarra:2012dw,Boddy:2015efa},
    \begin{equation}
      \frac{dN_\gamma}{dE} = \frac{4}{\Delta E} \Theta(E-E_-)\Theta(E_+-E)
    \end{equation}
    per annihilation. Here, $\Theta$ is the Heaviside step function,
    \begin{equation}
      E_\pm =
      \frac{m_\chi}{2}\left(1\pm\sqrt{1-\frac{m_{\pi^0}^2}{m_\chi^2}}\right)
    \end{equation}
    are the kinematic edges of the box, and $\Delta E \equiv E_+ - E_- =
    \sqrt{m_\chi^2 - m_{\pi_0}^2}$ denotes the box width.  

  \item $\chi\chi\rightarrow\pi^0\gamma$.
    This channel leads to a box and $\gamma$-ray line, with slightly different
    kinematics from the discussion above \cite{Boddy:2015efa}.

    The prompt photon spectrum per annihilation is
  \begin{equation}
      \frac{dN_\gamma}{dE} =  \delta(E-E_0) +\frac{2}{\Delta E}
      \Theta(E-E_-)\Theta(E_+-E)\;,
    \end{equation}
    where $E_0 = m_\chi - \frac{m_{\pi_0}^2}{4 m_\chi}$, $\Delta E = m_\chi -
    \frac{m_{\pi_0}^2}{4 m_\chi}$, and $E_\pm = \frac{m_\chi}{2}
    \left[\left(1+\frac{m_{\pi^0}^2}{4m_\chi^2}\right) \pm \left(1 -
    \frac{m_{\pi^0}^2}{4m_\chi^2}\right)\right]$.
\end{enumerate}

Finally, we briefly comment on the \emph{charged-pion channel}, that we neglect
otherwise in this work.  The dominant decay channel for charged pions is
$\pi^+\rightarrow \mu^+ \nu_\mu$ \cite{Olive:2016xmw}.  The muon will
subsequently decay into an electron and two neutrinos.  Therefore, the final
electron spectrum will receive two boosts, one from the annihilation and one
from the subsequent decay.  In addition, FSR can be produced in both the
annihilation as well as in the decays.  Because this channel does not add any
new spectral signatures, but rather smears out the ones studied in the other
channels, we ignore the charged pion channel in the present analysis.

%----------------------------
% Gamma-rays
%----------------------------

\subsection{Leptonic annihilation}

In the case of annihilation into leptons, we consider three benchmark channels
for the DM annihilation channels.  We comment here briefly on the role of
prompt photons for these channels and leave a detailed discussion about
secondary emission for the next section.
\begin{enumerate}[label=(\roman*)]
  \item
    $\chi\chi \rightarrow e^+e^-$.  In this case electrons and positrons are
    injected mono-energetically.  There is a significant contribution from
    final-state radiation (FSR), which we model following \cite{Beacom:2004pe}.
    An analytic expression is provided in appendix \ref{sec:functions}.
    
  \item
    $\chi\chi \rightarrow \phi \phi$, $\phi\rightarrow e^+e^-$.  Here, DM
    annihilates via a (scalar) mediator which subsequently decays into an
    electron/positron pair.  Now the positron spectrum will be boosted and has
    a box-like shape\footnote{Under the assumption that $m_\mathrm{e}\ll m_\phi
    < m_\chi$ and $\phi$ is a scalar.} \citep[e.g.~][]{Mardon:2009rc,
    Ibarra:2012dw}.  An analytic expression for FSR in such models is given in
    Eq.~(6) of \cite{Essig:2009jx} and for completeness repeated in appendix
    \ref{sec:functions}.

    FSR for this class of models is suppressed, enhancing the relative
    importance of secondary emission.  On the other hand, the final electron
    spectrum is boosted, softening any spectral features.  

  \item $\chi \chi \rightarrow \mu^+\mu^-$. 
    Like in the cascade channel, the injected electron spectrum gets boosted.
    On the other hand, this channel is accompanied by a large amount of FSR,
    since FSR can be produced at two stages: when dark matter annihilates to
    muons and in the subsequent decay of the muon to an electron and two
    neutrinos.  Both the positron and FSR spectra are obtained from
    \texttt{DarkSusy} \cite{Gondolo:2004sc}\footnote{P. Gondolo, J. Edsj{\"o},
    P. Ullio, L. Bergstr{\"o}m, M. Schelke, E.A. Baltz, T. Bringmann and G.
    Duda, http://www.darksusy.org}.
\end{enumerate}

Finally, we note that we do not consider the channel
$\chi\chi\rightarrow\tau^+\tau^-$, since the tau lepton is heavy,
$m_\tau=1.78\mathrm{\,GeV}$, and therefore only relevant for a very small
window in our considered energy range.

%-----------------------------------------------------------------------------------------
% S e c o n d a r y  F l u x e s
%-----------------------------------------------------------------------------------------

% %-----------------------------------------------------------------------------------------
% % Cosmic ray propagation
% %-----------------------------------------------------------------------------------------

\section{Gamma-ray signals from secondaries}
\label{sec:secondary}

\subsection{Radiative processes, cooling and timescales}

As mentioned above, we concentrate in this analysis on the annihilation signal
from the inner Galaxy (projected limits from dwarf spheroidal galaxies were
discussed in Ref.~\cite{Boddy:2015efa}).  The higher statistics from this
region makes it easier to identify spectral features, which are the main focus
of our work.  Furthermore, secondary emission components are in most cases
expected to be stronger.  However, in order to correctly predict $\gamma$-rays
from secondaries it is essential to understand all cooling and radiative
processes.  We begin this section with a brief discussion of all relevant
processes and timescales.

The relevant radiative processes in our energy window are bremsstrahlung,
inverse-Compton scattering (ICS) and synchrotron radiation.  The first two
yield X-ray and $\gamma$-ray photons, whereas the latter leads to radio
emission.  In addition, ionization and Coulomb losses are important
contributors to the overall energy losses.  Emissivities and energy-loss
timescales are well established and can be found in
Refs.~\cite{Blumenthal:1970gc, longair2011high, Evoli:2016xgn}.

In addition, we model in-flight annihilation (IfA) of positrons.  IfA results
from positrons colliding with ambient electrons.  The cross section for this
process is implemented following Refs.~\cite{Dirac:1930bga, Beacom:2005qv}.  An
expression for the differential cross section and resulting photon flux for a
single positron is given in appendix \ref{sec:functions}.  Note that at very
low energies, below those considered in this work, positrons can also
annihilate through the formation of positronium, a bound state of a positron
and an electron.  Positronium annihilation results in monochromatic photon
emission at 511 keV from the singlet (p-Ps) state and continuum emission from
the triplet (o-Ps) state.

\begin{figure}[t]
  \centering
  \includegraphics[width=\linewidth]{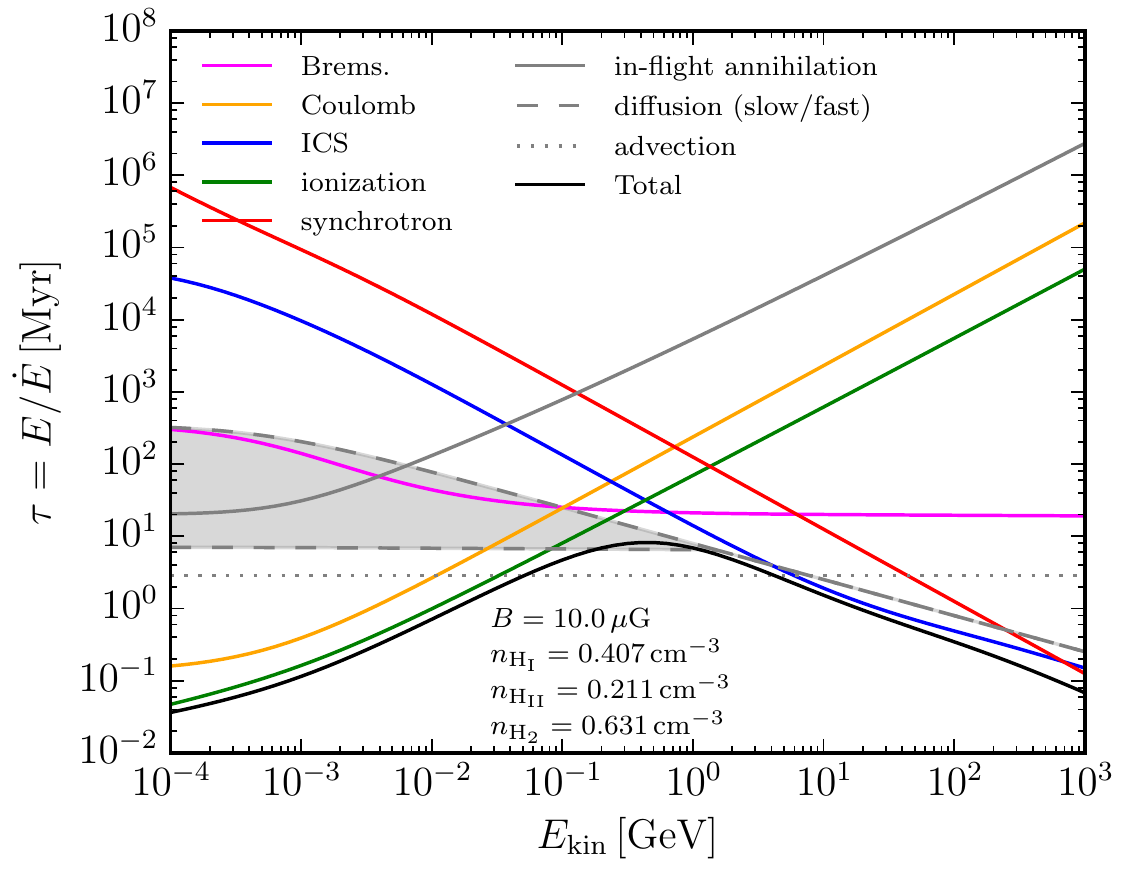}
  \caption{Timescales involved in propagation, radiative- and energy-loss
    processes.  Gas densities are averaged within a $\sim 0.7 \mathrm{\,kpc}$
    Galactocentric radius.  The advection and diffusion timescale corresponds
    to transport of $5^\circ$ ($\sim 0.7\mathrm{\,kpc}$) on the sky for a
    particle at the Galactic center.  The wind velocity is
    $v_\mathrm{wind}=250\mathrm{\,km\,s^{-1}}$.  The diffusion coefficient is
    given in Eq.~\ref{eq:diffusion} (slow/benchmark model).  The faster
    diffusion timescale at low energies is elaborated on in appendix
    \ref{sec:cr_transport}.}
\label{fig:timescales}
\end{figure}

In Fig.~\ref{fig:timescales} we show the energy-loss and annihilation
timescales for electrons and positrons from 1 MeV to 1 TeV in a medium
comparable to that of the Galactic bulge.  As can be seen, at energies
$\lesssim 100 \mathrm{\,MeV}$ IfA starts to dominate over ICS and synchrotron
losses.  Also bremsstrahlung becomes more important.  Energy losses are
dominated by ionization and Coulomb losses, with their relative importance
depending on the ionization fraction of the medium.  Note that bremsstrahlung,
IfA, ionization- and Coulomb losses all depend on the gas densities, whereas
ICS depends on the interstellar-radiation field and synchrotron emission on the
magnetic field.

%----------------------------
% Cosmic-Ray Propagation
%----------------------------
\subsection{Cosmic-ray propagation}
\label{CRprop}

We model the transport of CR electrons and positron in the Galaxy using the
numerical code \texttt{DRAGON}
\cite{Evoli:2008dv}.\footnote{www.dragonproject.org}.
\texttt{DRAGON} is designed to simulate all the relevant processes related to
Galactic CR propagation, in particular: diffusion, reacceleration, convection,
cooling (due to synchrotron, bremsstrahlung, inverse-Compton, Coulomb and
ionization), catastrophic losses (annihilation), and spallation.

The transport equation is solved for all CR species, from heavy nuclei down to
protons, antiprotons, and leptons. The code implements both the nuclear
spallation network taken from the public version of {\tt
  GALPROP}\footnote{{\tt GALPROP} project web page, http://galprop.stanford.edu} (see
e.g. \cite{Strong:2001fu} and references therein), and a complete cross-section
database obtained with the {\tt FLUKA} code \cite{2016APh....81...21M}.  The
code works in 2D and 3D mode; in the following we assume azimuthal symmetry and
work in the two-dimensional mode.

The annihilation and energy-loss rate is computed adopting the following
astrophysical ingredients:
\begin{itemize}
  \item The gas distribution is based on an azimuthally-averaged implementation
    of the detailed three-dimensional model described in
    Ref.~\cite{Ferriere:2007yq} for the Galactic bulge ($R < 3$ kpc), while the
    model presented in Ref.~\cite{Bronfman_1988} is used for the larger
    Galactocentric radii. Within $R\lesssim 0.7$ kpc the ionization fraction is
    $n_\mathrm{H^+} / (n_\mathrm{H} + 2n_\mathrm{H_2}) \approx 0.1$.
  \item The magnetic field is taken from \cite{2011ApJ...738..192P} and is
    derived by a wide set of Faraday Rotation measurements.
  \item The interstellar radiation field is taken from the public version of
    {\tt GALPROP}; it is described in \cite{Porter2005, Porter2008}.
\end{itemize}

Diffusion of sub-GeV electrons remains largely unconstrained (partially due to
the lack of observational data).  We assume here isotropic diffusion with
power-law dependence on momentum, $p$.  The diffusion coefficient is
\begin{equation}
	\label{eq:diffusion}
	D(p) = \beta D_0 \left(\frac{p}{p_0}\right)^\delta\;,
\end{equation}
where $\beta = v/c$, $p_0=4\mathrm{\,GeV}$, $\delta=\frac 1 2$ and $D_0 =
4\times 10^{28}\mathrm{\,cm^{2}\,s^{-1}}$.  We verified with {\tt DRAGON} that
these parameters are compatible with the current AMS-02 measurements of the
boron-over-carbon ratio in the GeV-TeV range \cite{PhysRevLett.117.231102}.  We
comment on the effects of changing these assumptions on the diffusion
coefficient in Sec.~\ref{sec:discussion}.  For simplicity, we do not consider
diffusive reacceleration in this work.

Soft X-rays provide evidence for a \emph{galactic wind} \cite{Everett:2007dw}.
Following Ref.~\cite{Everett:2007dw} we implement a cylindrical Galactic wind
of radius $R_\mathrm{wind}=3\mathrm{\,kpc}$ and velocity
$v_\mathrm{wind}=250\mathrm{\,km\,s^{-1}}$ centered on the Galactic center into
our benchmark model.  As an alternative we also show results in
absence of a wind in appendix \ref{sec:cr_transport}. 

The relevant timescales for diffusion and advection are included in
Fig.~\ref{fig:timescales}.  Under the above benchmark assumptions, advection
dominates over diffusion below $\sim 10\mathrm{\,GeV}$.  Moreover, cooling is
much faster than diffusion. In the absence of a wind, sub-GeV particles would
approximately cool in-situ. However, when a wind is present
($v_\mathrm{wind}=250\mathrm{\,km\,s^{-1}}$), a particle at the Galactic Center
(GC) will be advected over $5^\circ$ on the sky. We apply a region-of-interest
of $10^\circ\times10^\circ$ centered at the GC, implying that in the presence
of a wind a particle is more likely to be advected out of our ROI than to
radiate its energy.

\subsection{Secondary gamma-rays}

\begin{figure}[t]
  \begin{center}
        \includegraphics[width=\linewidth]{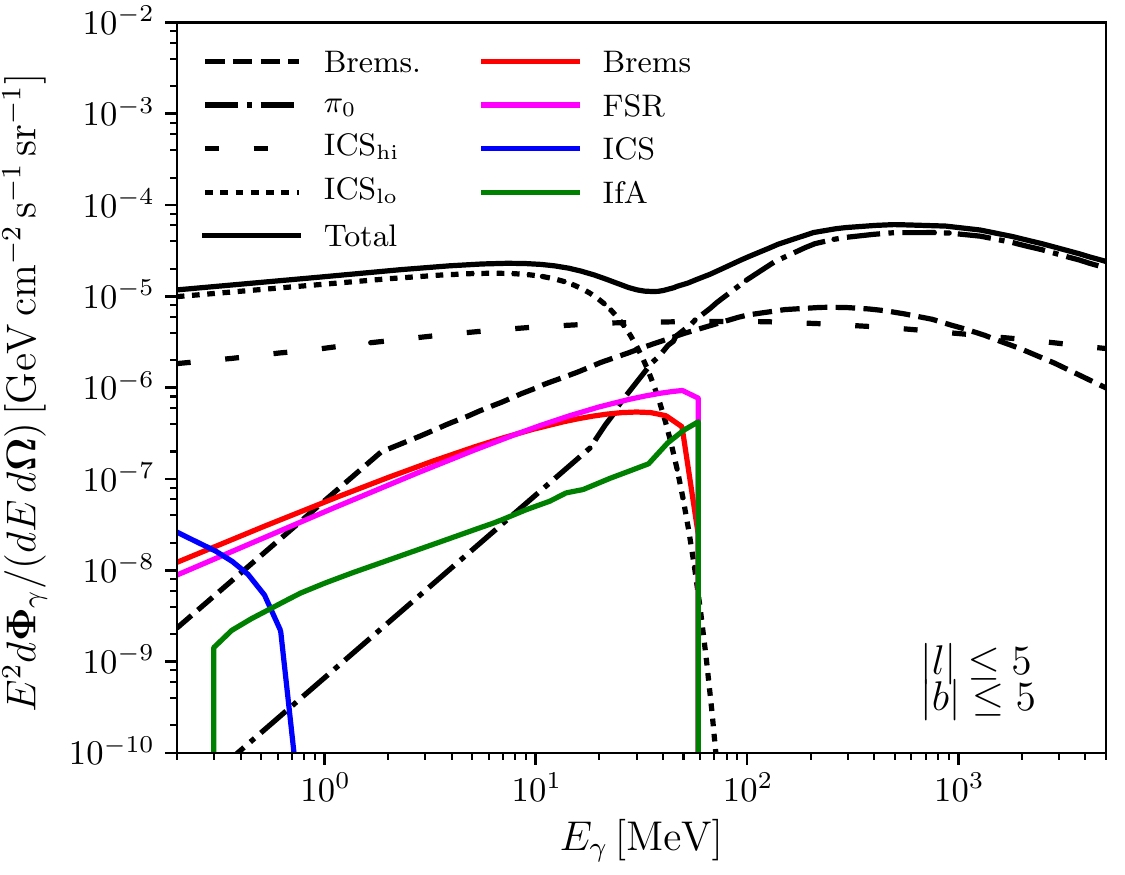}
  \end{center}
  \caption{$\gamma$-ray spectrum resulting from $\chi\chi \rightarrow e^+e^-$
  with $\left<\sigma v\right>=10^{-28}\mathrm{\,cm^3\,s^{-1}}$ in the inner
  $10^\circ\times 10^\circ$ of the Galaxy.  The DM signal is broken up into
  individual components: IfA (green), FSR (magenta), bremsstrahlung (red) and
  ICS (blue).  Black lines indicate the various diffuse-background components.  }
  \label{fig:spectrum}
\end{figure}

Next, the $\gamma$-ray spectrum is obtained by convolving the steady-state
electron and positron spectra with either the gas-distribution (bremsstrahlung
and IfA) or the interstellar-radiation field (ICS) and performing an integral
over the line-of-sight.  We ignore emission from synchrotron radiation, since
it will be in radio, but we do include it as a cooling term.  In
Fig.~\ref{fig:spectrum} we show an example $\gamma$-ray spectrum for a
$m_\chi=60\mathrm{\,MeV}$ DM particle annihilating to $\chi\chi \rightarrow
e^+e^-$.  Note that there is a kinematic cutoff in the IfA spectrum at
$E_\gamma= m_e/2$ \cite{1969Ap&SS...3..579S} (for more details see appendix
\ref{sec:functions}).

\smallskip

%-----------------------------------------------------------------------------------------
% S E N S I T I V I T Y  C A L C U L A T I O N
%-----------------------------------------------------------------------------------------
\section{Background modeling and sensitivity projections}
\label{sec:sensitivity}

%----------------------------
% e-ASTROGAM
%----------------------------
\subsection{Instrumental details}

\begin{table}[h]
    \centering
    \begin{tabular}{cccc}
        \toprule
	\hline
    \hline
    \multicolumn{4}{c}{\bf{Compton domain}}\\
    & e-ASTROGAM & COMPAIR & ADEPT \\
	\hline
    Energy range $\mathrm{[MeV]}$		& 0.3--10		& 0.2 -- 10 	& -- \\
    $\Delta E / E$						& 1.3\%			& 2\%--5\%		& -- \\
	$A_\mathrm{eff}\mathrm{\,[cm^2]}$	& 50--560		& 50 -- 250		& -- \\
    FoV $\mathrm{[sr]}$					& 2.9			& 3				& -- \\
	\hline
    \hline
    \multicolumn{4}{c}{\bf{Pair-conversion domain}}\\    
    & e-ASTROGAM & COMPAIR & ADEPT \\
    \hline
    Energy range $\mathrm{[MeV]}$		& 10--3000		& 10--500	 	& 5--200 	\\
    $\Delta E / E$						& 20--30\%		& 12\%			& 30\% 		\\
	$A_\mathrm{eff}\mathrm{\,[cm^2]}$	& 215--1810		& 20--1200		& 50--700	\\
    FoV $\mathrm{[sr]}$					& 2.5			& 3				& 3 		\\
	\hline
    \hline
	\end{tabular}
    \caption{Instrumental details for three proposed $\gamma$--ray
      telescopes, e-ASTROGAM \cite{DeAngelis:2016slk}, COMPAIR
      \cite{Moiseev:2015lva} and ADEPT \cite{Hunter:2013wla}.  In this work we
      assume an on-target observation time of $T_\mathrm{obs}=1\mathrm{\,yr}$.
      Our reference experimental scenario is e-ASTROGAM performing a full-sky
      survey. All details are taken from Ref.~\cite{DeAngelis:2016slk}. The
      energy resolution in the pair conversion regime is set to $\Delta E / E =
      30\%$.
    }
    \label{tab:instrument}
\end{table}

We assume a future instrument similar to the mission concept e-ASTROGAM
\cite{DeAngelis:2016slk}.  Marginal differences put aside, our results also
roughly apply to other proposed mission such as AMEGO, COMPAIR \cite{Moiseev:2015lva}
or ADEPT \cite{Hunter:2013wla} within their energy range.  An important reason
for this is that our projected sensitivity turns out the be systematics limited
over most of the considered DM mass range.  This is a consequence of our use of
a relatively large ROI on the inner Galaxy.  This makes slight variations in
the effective area irrelevant for the projected limits. On the other hand, a
better spectral resolution, such as that of COMPAIR, could potentially improve
the projected limits in the pair--regime by a factor two (see 
Tab.~\ref{tab:instrument}).  We provide an overview of instrumental details and
compare instruments in Tab.~\ref{tab:instrument}. These instruments can fill
the MeV gap and as we will show have great prospects to hunt for MeV DM. At low
energies ($<10\mathrm{\, MeV}$) the detection principle is based on Compton
scattering (similar to COMPTEL) and at higher energies on the detection of
$\gamma$-rays via pair-conversion (like \Fermi-LAT).

All instrumental details applied in our analysis are taken from Fig.~17 and
tables 2 and 3 of Ref.~\cite{DeAngelis:2016slk}.  The effective area
($A_\mathrm{eff}(E)$) is modeled as a function of energy and we interpolate
between the reference values.  It is typically of the order $A_\mathrm{eff}\sim
\mathrm{few}\times 100 (1000)\rm\,cm^2$ in the Compton (pair-production)
domain.  The spectral resolution is modeled as a Gaussian with energy dependent
width. It is of the order $\Delta E / E \sim 0.01 (0.3)$, in the Compton
(pair-production) domain.  The excellent energy resolution in the Compton
domain yields great power to search for sharp spectral features such as \gr
lines, as we will demonstrate below (Fig.~\ref{fig:limits_pions}).  Finally, we
adopt an effective observation time on our ROI of
$T_\mathrm{obs}=1\mathrm{\,yr}$.  Considering that the field-of-view of
e-ASTROGAM is $2.5\mathrm{\,sr}$, this corresponds to $\sim5$ years of pure
observation time with uniform sky coverage.

%----------------------------
% Background components
%----------------------------
\subsection{Background components}

Our diffuse background model consists of four components as shown by the black
lines in Fig.~\ref{fig:spectrum}.  Three of them -- the bremsstrahlung, $\pi^0$
and inverse-Compton components -- characterize the diffuse background at higher
energies.  We take the spectral templates from \cite{Strong:2011pa} (see also
\cite{Bouchet:2011fn}). Those models were computed with the \texttt{GALPROP}
code \cite{Strong:1998pw}, and fitted to data in the window $|l|<30^\circ,
|b|<10^\circ$.  Since the region of interest (ROI) for this analysis is
$|l|<5^\circ, |b|<5^\circ$, we rescale those models to match our ROI, assuming
the same morphology computed with \texttt{DRAGON} at 1 GeV.  We checked that
the overall \gr intensity of our background model in the ROI is compatible with
\Fermi-LAT data.

The fourth component characterizes the background at lower energies and is
modeled as a power-law, following \cite{Beacom:2005qv}, and including a
super-exponential cutoff:
\begin{equation}
  \label{eq:ICSlo}
  \frac{d\Phi}{dE}=0.013\left(\frac{E}{1\mathrm{\,MeV}}\right)^{-1.8}
  e^{-\left(\frac{E}{2\mathrm{\,MeV}}\right)^2}
  \mathrm{\,cm^{-2}\,s^{-1}\,sr^{-1}\,MeV^{-1}}.
\end{equation}
This component, possibly originating from Inverse-Compton scattering, is
introduced to get a reasonable agreement with the COMPTEL data in the region
$|l|<30^\circ, |b|<5^\circ$ \cite{Beacom:2005qv}.

We will henceforth refer to this emission as $\mathrm{ICS_{lo}}$ and to the
inverse-Compton model at higher energies as $\mathrm{ICS_{hi}}$.  Similarly to
the other three background models, we map the intensity to that in our ROI
utilizing the \texttt{DRAGON} ICS template.

Our full background model is then described by
\begin{equation}
  \label{eq:background}
  \phi_\mathrm{bg}=\sum_{i=1}^4 \theta_i \phi_i.
\end{equation}
Here $\phi(E, \Omega) = d \Phi /dE d\Omega$ and $i=1,2,3,4$ refers to
bremsstrahlung, $\mathrm{ICS_{hi}}$, $\pi^0$ and $\mathrm{ICS_{lo}}$,
respectively. The parameters $\theta_i$ are the normalization of the various
components. The baseline background model has $\theta_i=1$ for components
$i=1,2,3$ and for component $i=4$ such that the latter reproduces
Eq.~\ref{eq:ICSlo} within $|l|<30^\circ, |b|<5^\circ$

It is important to emphasize that -- for the purposes of this paper -- an
approximate treatment of the background suffices.  A dedicated, more accurate
modeling of the diffuse \gr sky from MeV to GeV and a corresponding, consistent
modeling of the dark matter signal is well beyond the scope of the present
paper and will be left for future work.

%----------------------------
% FISHER information
%----------------------------

\subsection{Projected limits from Fisher forecasting with correlated background
systematics}

We compute here projected 95\% confidence level (CL) upper limits on any DM
signal, assuming that no signal is present in the data.  Fisher forecasting is
applied to calculate projected limits for a mission similar to e-ASTROGAM~\cite{Edwards:2017mnf}.

We consider a signal spectrum given by the differential flux
$\phi(E,\Omega)=d\Phi/dE(\Omega)$ in units
$\mathrm{ph\,cm^{-2}\,s^{-1}\,GeV^{-1}}$, composed of various additive
components with subscript $i$,
\begin{equation}
  \phi_S(E, \Omega|\vec \theta_S) = \sum_i \theta_{S,i}\ \phi_{S,i}\;.
\end{equation}
In our case, we include the components
$i\in\{\mathrm{bremsstrahlung,\,FSR,\,ICS,\,IfA}\}$.  

Additionally, we consider the background and foreground components
\begin{equation}
  \phi_{bg}(E, \Omega|\vec\theta_{bg}) = \sum_i\theta_{\mathrm{bg}, i}\phi_{\mathrm{bg},i}\;,
\end{equation}
as described in Eq.~\ref{eq:background}.  The total emission is then given by
$\phi = \phi_S + \phi_\mathrm{bg}$.

\medskip

The Fisher information matrix is an $N\times N$ matrix where $N$ is the number
of parameters $\vec\theta$.  Given a likelihood function, $\mathcal{L}(\vec
\theta|\mathcal D)$, it is defined as
\begin{equation}
  \mathcal{I}_{ij}(\vec \theta) = 
  	-\left< \frac{\partial^2 \ln \mathcal L (\vec \theta|\mathcal D)}{
    \partial \theta_i \partial\theta_j}\right>_{\mathcal D(\vec \theta)},
\end{equation}
where the average is taken over multiple realizations of the data, $\mathcal
D$.  Practically, the average is taken over
$\mathcal{D}(\vec\theta_\mathrm{S}=\vec 0, \vec\theta_\mathrm{bg}=\vec 1)$ when
setting upper limits.

From here on we will assume an unbinned Poisson likelihood for the description
of mock data.  Additionally, we include an effective model for correlated
instrumental systematics and/or background model uncertainties in our analysis,
by making the substitution
\begin{equation}
  \phi_\text{bg} \to (1+\delta(E)) \phi_\text{bg}\;.
\end{equation}
Here, $\delta(E)$ parametrizes general fractional deviations from the nominal
background model, which are assumed to be correlated over smaller and larger
energy ranges, as specified below.  One can then show that the Fisher
information becomes~\cite{Edwards:2017mnf}
\begin{align}
  \label{eq:Itot}
  \mathcal{I}_{ij} = \sum_{ab}
  \frac{\partial_i\phi}{\phi_\mathrm{bg}}(E_a)
  D_{ab}^{-1}
  \frac{\partial_j\phi}{\phi_\mathrm{bg}}(E_b)
  \;,
\end{align}
with $\partial_i\phi \equiv d\phi/d\theta_i$.  Here, $E_a$ (and equivalently
$E_b$) refer to a dense grid of reference energies at which $\delta(E)$ and all
other fluxes are evaluated.  It has to be sufficiently dense to capture all
spectral variations.  For this analysis we use 2000 logarithmically spaced bins
between $0.5 \mathrm{\,MeV}$ and $5 \mathrm{\,GeV}$. We note that the secondary
\gr~ spectra obtained through \texttt{DRAGON} have a more coarse binning due to
computational constraints.  In practice we interpolate over these spectra to
get a sufficiently fine binning.

Furthermore, $D_{ab}$ is defined as
\begin{align}
  D_{ab} \equiv \frac{\delta_{ab}}{\Delta E_a \mathcal{E}(E_a)\phi_\mathrm{bg}(E_a)} + 
  \Sigma_\text{bg}(E_a, E_b)\;,
\end{align}
where $\Delta E_a$ is the energy step between two consecutive values of $E_a$.
Furthermore, $\mathcal{E}(E)\equiv T_\mathrm{obs}A_\mathrm{eff}(E)$ denotes the
exposure,\footnote{In principle, the exposure is a function of not only energy,
but also sky-position through $T_\mathrm{obs}(\Omega)$.  However, for this
paper we assume uniform sky coverage.} which is an energy dependent quantity.
Finally, $\Sigma_\text{bg}$ denotes the covariance matrix that describes the
background model uncertainties encoded in $\delta(E)$ (not to be confused with
the covariance matrix of the model parameters, which equals
$\mathcal{I}^{-1}$).  It can be thought of as the covariance matrix of a
Gaussian random field, with $\Sigma_\text{bg}(E, E') = \langle
\delta(E)\delta(E')\rangle$, where the average is taken over many realizations
of the function $\delta(E)$.  It is here parameterized as
\begin{align}
  \Sigma_\text{bg}(E, E') = \sum_{k=1,2}\sigma_\mathrm{syst}^{(k)}(E) \sigma_\mathrm{syst}^{(k)}(E') \rho_k(E, E'),
\end{align}
where we assumed two independent contributions.  Here
$\sigma^{(k)}_\mathrm{syst}(E)$ is the overall magnitude of the systematic
uncertainty at energy $E$, while $\rho_k(E, E')$ parametrizes the correlation
between systematics at different energies and equals one along the
diagonal.  We adopt the simple form (motivated by a log-normal distribution)
\begin{align}
  \rho_k(E, E') = e^{-\frac12 \left(\frac{\ln E/E'}{w_k}\right)^2}\;.
  \label{eqn:rho}
\end{align}
As mentioned above, we use here two independent covariance matrices, which are
added ($k=1,2$), and which correspond to background systematic with short and
long correlation scales in energy space.  We adopt 2\%
($\sigma^{(1)}_\text{syst}=0.02$) as the short-scale correlated systematic,
with a correlation length of just $1\%$ ($w_1=0.01$) in energy space, which is
motivated by the results from Refs.~\cite{Albert:2014hwa, Ackermann:2015lka} and presumably mostly
of (unknown) instrumental origin. For systematics correlated over a larger
energy range, possibly related to diffuse emission modeling and uncertainties
in the effective area, we assume a value of 15\%
($\sigma_\text{syst}^{(2)}=0.15$), with the correlation length taken as 0.5 dex
($w_2=1.15$).  This is reasonably representative of what is found when
analyzing Galactic diffuse emission along the Galactic
disk~\cite{Calore:2014xka}.

We note that our treatment of systematic uncertainties for future instruments
remains necessarily uncertain, although we believe that our choices are
plausible.  However, we checked that our qualitative results are not sensitive
to changes in in these parameters, and the quantitative results are fairly
independent of $w$ and behave as expected for changes in $\sigma$. For a more
elaborate discussion see Sec.~\ref{sec:discussion}.

When systematic uncertainties are absent, $\Sigma_\text{bg}\rightarrow 0$, the
Fisher information can be written in the more common form in terms of the
signal-to-noise ratio \cite{Edwards:2017mnf},
\begin{align}
  \label{eq:Itot_original}
  \mathcal{I}_{ij}=\int_\mathrm{E_{min}}^\mathrm{E_{max}}dE\int_\mathrm{ROI} d\Omega\,
  \mathcal{E}(E) \frac{\partial_i\phi(E, \Omega) \partial_j\phi(E, \Omega)}
  {\phi_\mathrm{bg}(E, \Omega|\vec\theta=\vec{1})}\;.
\end{align}

\bigskip

Finally, in the background-limited regime we can construct an upper limit of
$100\cdot(1-p)\%$ confidence level (CL)  on a signal parameter,
$\theta^{UL}_{S,i}$, by inverting the Fisher-information,
\begin{align}
  \theta_{S,i}^{UL} = Z(p) \sqrt{\mathcal{I}^{-1}_{ii}}\;,
\end{align}
where $Z(p) = \mathrm{ppf}(1-p)$, with $\mathrm{ppf}(x)$ the inverse of the
cumulative-distribution functions, also known as the percent-point function.
For 95\% CL (or $p=0.05$) $Z(p)=1.645$.

%---------------------------------------------------------------------------------------
% Results
%---------------------------------------------------------------------------------------

\section{Results}
\label{sec:results}

%----------------------------
% Limits for gamma-ray lines and pions
%----------------------------

\subsection{Projected limits for $\gamma$-ray lines and pions}

\begin{figure}[t]
     \begin{center}
        \includegraphics[width=\linewidth]{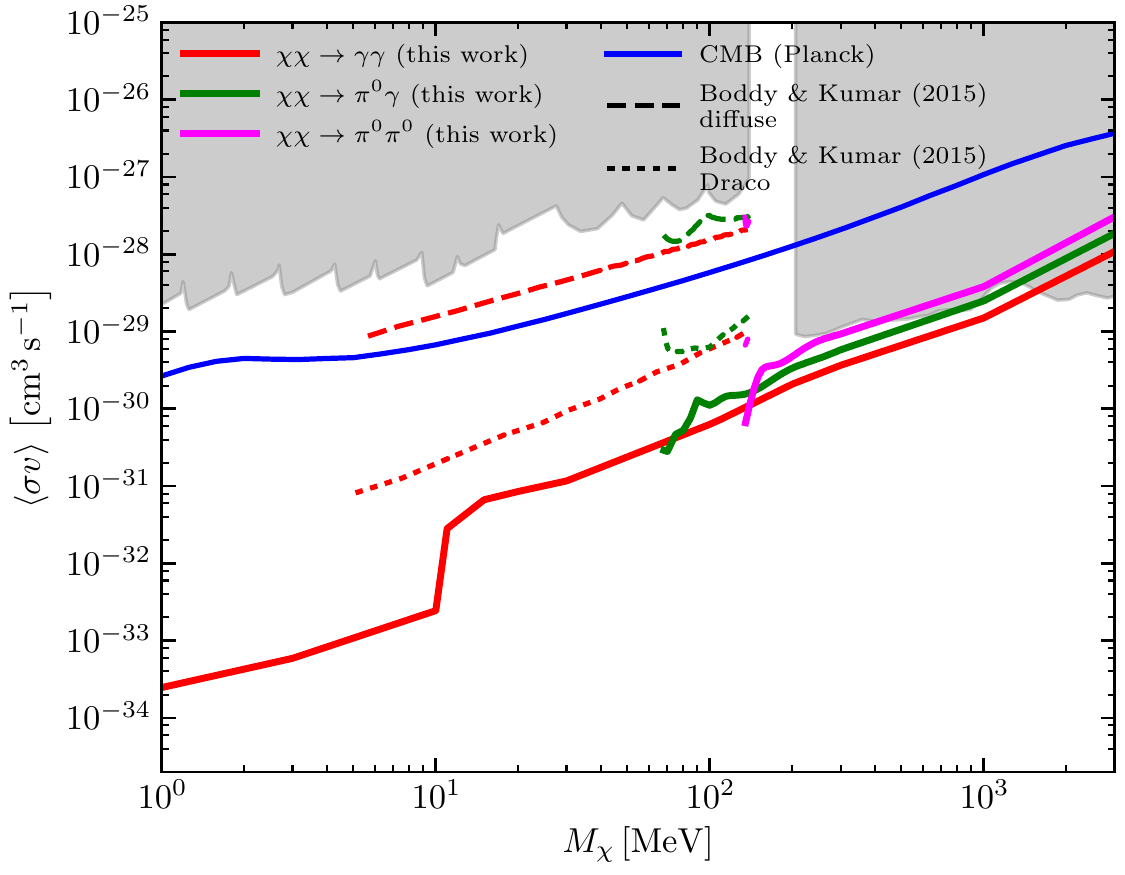}
    \end{center}
    \caption{Projected 95\% CL upper-limit on DM annihilating to $\gamma\gamma$
      (red), $\pi^0\gamma$ (green) and $\pi^0\pi^0$ (magenta).  Projected
      limits from \cite{Boddy:2015efa} are shown in the same colors for diffuse
      emission (dashed) and for the Draco dwarf-spheroidal galaxy (dotted).
      The CMB constraints from
    Planck are shown as a solid blue line, they are for
  $\chi\chi\rightarrow\gamma\gamma$ \cite{Ade:2015xua, Slatyer:2015jla}.  For
  the same channel we show the limits derived by \cite{Boddy:2015efa} from
  COMPTEL ($|b|>30^\circ$), EGRET ($20^\circ<|b|<60^\circ$) and \Fermi--LAT
  ($|b|>20^\circ$) from diffuse \grs as the shaded light-grey area.
  Additionally, there are \Fermi--LAT inner-Galaxy limits on monochromatic
\grs~ for DM with $m_\chi>200\mathrm{\,MeV}$ \cite{Ackermann:2015lka}.}
    \label{fig:limits_pions}
\end{figure}

Figure \ref{fig:limits_pions} shows the projected upper limits from the
Galactic center for annihilation into mono-chromatic photons (red), a neutral
pion and a photon (green) and two neutral pions (magenta).  All projected
limits are 95\% CL, for one year of effective exposure and for instrumental
specifications similar to those of e-ASTROGAM.  The emission is prompt and
therefore traces exactly the DM distribution.

The limits from monochromatic photons are only slightly stronger than those
from box-spectra due to pion decay. This is a consequence of the spectral
resolution in the pair-production domain of e-ASTROGAM, assumed to be
$\sim30\%$, which significantly broadens the line feature: as a result, the
line is not much sharper than the box-like feature.  The fact that pion decay
yields two photons, leading to twice as many photon in $\chi\chi\rightarrow
\pi^0\pi^0$ compared to direct annihilation into photons, further reduces the
difference.  Below 10 MeV the limits for monochromatic photons improve by an
order of magnitude, because of the better spectral resolution in the Compton
domain (a factor $\sim 10$ better with respect to the pair--creation regime).

A MeV mission similar to the proposed e-ASTROGAM could outperform CMB
constraints for $s$-wave annihilating DM (shown for
$\chi\chi\rightarrow\gamma\gamma$ in Fig.~\ref{fig:limits_pions}) by more than
one order of magnitude below 1 GeV.

\medskip

Our projections are compared to \Fermi--LAT limits from the inner-Galaxy on
monochromatic photons resulting from DM annihilation with $m_\chi>200
\mathrm{\,MeV}$ as obtained by \cite{Ackermann:2015lka}. The limits portrayed
correspond to the analysis optimized for a NFW density profile.

In addition, we compare our limits to the existing and forecasted limits from
Ref.~\cite{Boddy:2015efa}.  The \emph{existing} limits, shown as the light-grey
shaded regions, are derived from the diffuse \gr flux measured by COMPTEL
($|b|>30^\circ$), EGRET ($20^\circ<|b|<60^\circ$) and \Fermi--LAT
($|b|>20^\circ$).  Our projected limits for the diffuse \gr sky suggest that
these constraints can be improved by at least two orders of magnitudes.  We
note that a reanalysis of existing data and modeling of background could
improve the current situation already now significantly, and would be an
interesting endeavour.

For the \emph{forecast}, Ref.~\cite{Boddy:2015efa} assumes an ADEPT--like
instrument with a spectral resolution of $\Delta E / E = 15\%$ (a factor two
better than what we assume), an effective area of
$A_\mathrm{eff}=600\mathrm{\,cm^2}$ and a systematic uncertainty in the
backgrounds of 15\%.  The total observation time of the experiment is set to 5
years.  Ref.~\cite{Boddy:2015efa} considers two targets, and the corresponding
projections are shown in Fig.~\ref{fig:limits_pions}: the optimistic projection
for diffuse \grs above $|b|>30^\circ$ (dashed) and for the Draco
dwarf-spheroidal galaxy (dotted).  A detailed discussion of the differences
will be done in Sec.~\ref{sec:discussion}.

%----------------------------
% Limits for leptonic channels
%----------------------------
\subsection{Projected limits for leptonic channels}
\begin{figure}[t]
     \begin{center}
        \includegraphics[width=\linewidth]{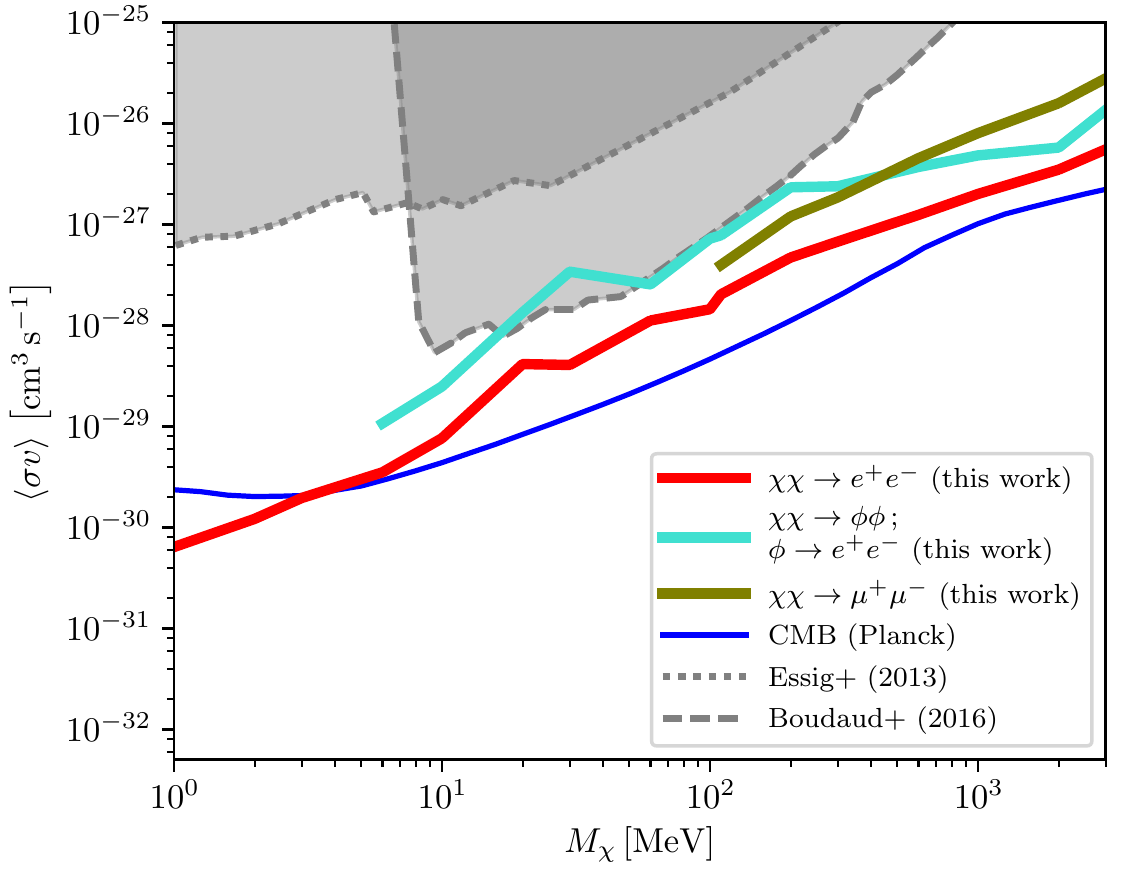}
    \end{center}
    \caption{
    Projected 95\% CL upper-limit on $\gamma$-ray
    emission from DM annihilating to $\mathrm{e^{+}e^{-}}$.
    Results are for the total DM spectrum from the
    three reference leptonic cases: direct annihilation (red),
    cascade channel (turquoise) and the muon channel (olive).
    The blue solid line shows the CMB limits on
    DM $s$-wave annihilation into $e^+e^-$
    from Planck for $s$-wave annihilating DM \cite{Ade:2015xua, Slatyer:2015jla}.
    In addition we show in light-grey the limits for $\chi\chi\rightarrow e^+e^-$ 
    from Voyager (dashed) \cite{Boudaud:2016mos} and current 
    limits from diffuse emission (dotted) \grs~\cite{Essig:2013goa}.
    }
    \label{fig:limits_leptons}
\end{figure}

In Fig.~\ref{fig:limits_leptons} we show the spectral constraints (95\% CL)
obtained for our three reference leptonic channels.  We adopt the transport
model described in section \ref{CRprop}: The dominant effect in the low-energy
range considered in this work is advection, caused by a Galactic wind modeled
with $v_\mathrm{wind}=250\mathrm{\,km\,s^{-1}}$ (see also
Fig.~\ref{fig:timescales}).  The impact of diffusion is negligible at low
energies for our benchmark scenario, in which we consider a power-law
extrapolation of the diffusion coefficient tuned on GeV boron-over-carbon AMS
data.

\begin{figure}[t]
     \begin{center}
        \includegraphics[width=\linewidth]{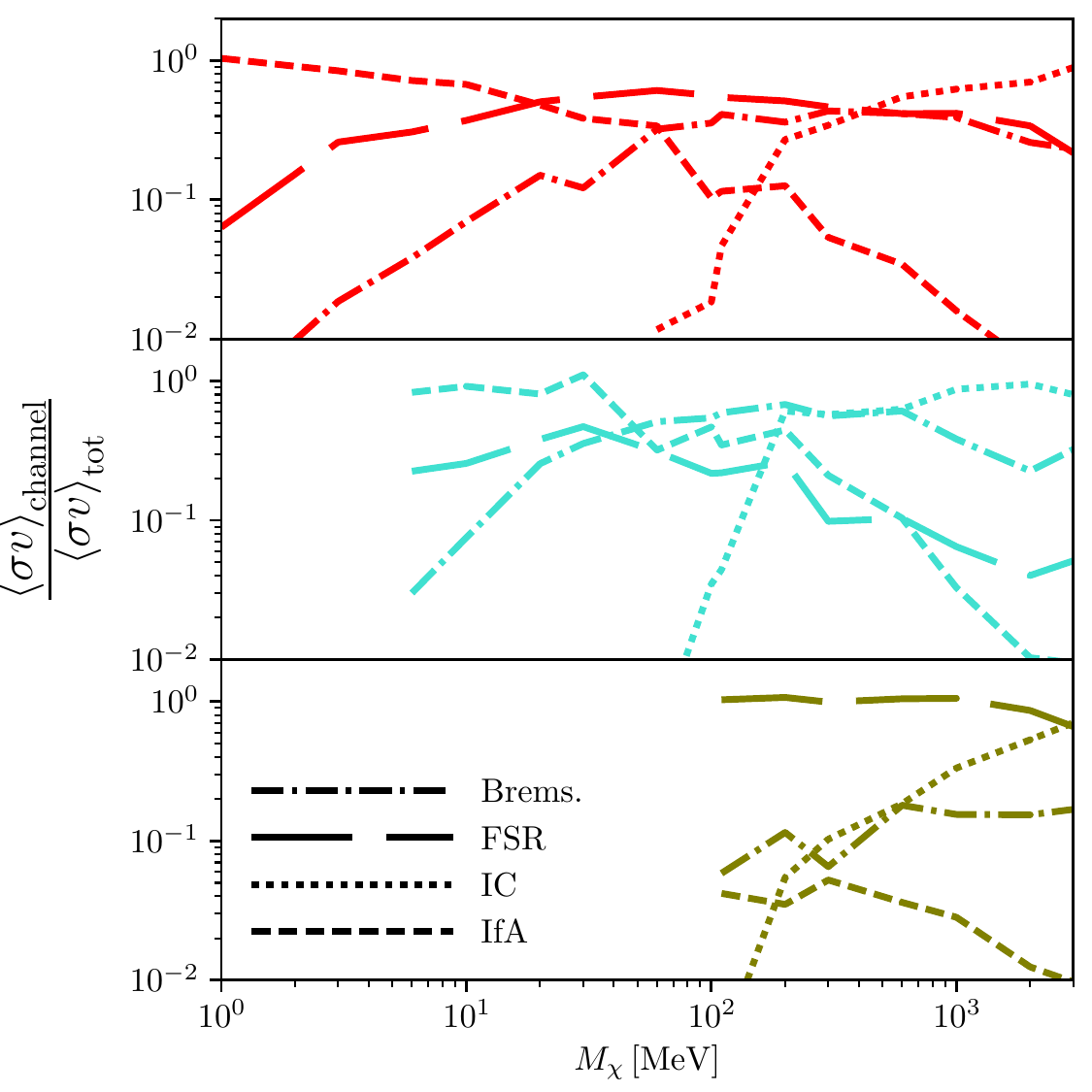}
    \end{center}
    \caption{
    Ratio of the limit obtainable from a single emission
    component (i.e., bremsstrahlung, FSR, ICS or IfA) to that of the full
    DM spectrum. The panels, from top to bottom, are for
    $\chi\chi\rightarrow e^{+}e^{-}$,
    $\chi\chi\rightarrow \phi\phi \rightarrow e^{+}e^{-}e^{+}e^{-}$ and
    $\chi\chi\rightarrow \mu^{+}\mu^{-}$.
    In case of annihilation to $e^+e^-$ (cascade annihilation)
    secondary emission in the form of in-flight annihilation 
    contributes most to the limits below $\sim20\,(50)\mathrm{\,MeV}$
    and offers the best channel for detection.}
    \label{fig:limits_leptons2}
\end{figure}

The three panels in Fig.~\ref{fig:limits_leptons2} show the ratio of the limit
obtained from a single emission component (i.e., bremsstrahlung, FSR, ICS or
IfA) over that of the full DM spectrum for the electron/positron, cascade and
muon channel, respectively.  The plots show the importance of each emission
component.  In case of mono-energetic injection of electron-positron pairs,
in-flight annihilation of positrons dominates the bounds below $\sim 20$ MeV.
For the cascade annihilation scenario the upper limits arise predominantly from
IfA below $\sim 50$ MeV, since FSR is suppressed. However, the overall limit is
somewhat weaker due to the softening of the injected lepton spectrum (see
Fig.~\ref{fig:spec_cascade} in appendix \ref{sec:spectral_plots}).  From $\sim
50$--$200$ MeV bremsstrahlung provides the dominant signal. The muonic channel
is most easily detectable through FSR at all DM masses.  In this case FSR
arises at two stages, when DM annihilates to muons, and in the subsequent decay
of the muon.  ICS dominates the bounds above a few hundred MeV for the direct
channel into $e^+e^-$ and for the cascade channel. However, unlike FSR, IfA and
to some extent bremsstrahlung the ICS spectrum is not very peaked and will
therefore be more difficult to distinguish from any astrophysical background.

In light-grey we show the recent limits from Voyager on $\chi\chi\rightarrow
e^+e^-$ (Fig.~\ref{fig:limits_leptons} \cite{Boudaud:2016mos}). Future diffuse
\gr studies can surpass these limits for MeV DM and cover a broader mass region
in general. In addition, we show existing diffuse \gr limits
\cite{Essig:2013goa} (based on INTEGRAL and COMPTEL data, and without
performing any background subtraction).  A future dedicated MeV \gr experiment
can improve these limits by two to three orders-of-magnitude (less impressive
but still interesting improvements could be obtained by a dedicated reanalysis
of available data, including a proper modeling of backgrounds).

Figure ~\ref{fig:limits_leptons} shows that the CMB limits from Planck (blue
solid line) for $\chi\chi\rightarrow e^+e^-$ through an $s$-wave process are
stronger than what is attainable with a MeV mission like e-ASTROGAM.  However,
as mentioned above thermal relic models will require $s$-wave suppression in
order to evade the CMB bounds. In this case the CMB constraints disappear, whereas
the limits from diffuse \grs remain.

%----------------------------
% Other constraints
%----------------------------
\subsection{Other constraints}
\label{sec:constraints}
We here briefly review limits from the CMB, BBN and the $p$--wave thermal relic
cross section to which we compare our results for the projected upper limits
from future diffuse \gr studies.

Regarding the CMB bounds for late time energy injection from Planck, these
apply to $s$--wave annihilation.  The constraint on the so-called annihilation
parameter, $p_\mathrm{ann}$ is (for the Planck TT, TE, EE+lowP data)
\cite{Ade:2015xua}:
\begin{equation}
  p_\mathrm{ann}\equiv\frac{f_\mathrm{eff} \left<\sigma v\right>}{m_\chi}< 4.1\times10^{-31}\mathrm{\,cm^3\,s^{-1}\,MeV^{-1}}\;.
\end{equation}
Here $f_\mathrm{eff}$ is an efficiency parameter that relates the total
injected energy to the energy that is used to increase the ionization fraction.
We adopt values for $f_\mathrm{eff}$ for $\chi\chi\rightarrow \gamma \gamma$
and $\chi\chi\rightarrow e^+e^-$ from
Ref.~\cite{Slatyer:2015jla}\footnote{Tables available from
\url{http://nebel.rc.fas.harvard.edu/epsilon.}}. 
Constraints for the cascade channel, which are not shown, 
are only marginally different from
those for the direct annihilation channel \cite{Elor:2015bho}.
For the muonic channel they are weaker by a factor of a few 
\cite{Slatyer:2015jla,Elor:2015bho}.
For a future experiment that
is cosmic-variance (CV) limited the bound can be as strong as $p_\mathrm{ann}<
8.9\times10^{-32}\mathrm{\,cm^3\,s^{-1}\,MeV^{-1}}$, about five times stronger
than the current Planck bound\cite{Galli:2009zc}.  CMB limits from Planck for
late time energy injection are displayed as a blue solid line in
Figs~\ref{fig:limits_pions} and \ref{fig:limits_leptons}.  We note that any
freeze-out thermal relic DM candidate will require velocity suppressed
annihilation in order to evade the CMB bounds.  For $p$-wave annihilation the
CMB bounds deteriorate by a few orders of magnitude \cite{Diamanti:2013bia}.

In addition, for low mass thermal dark matter, there exist constraints from
big-bang nucleosynthesis (BBN) and from the effective number of neutrinos as
inferred from the CMB by Planck. The latter provide the stronger constraints,
ruling out a thermal dark matter candidate in the form of a Majorana (Dirac)
fermion  below $m_\chi < 3.5\, (7.3) \mathrm{\,MeV}$, independent of whether
the annihilation is $s$- or $p$-wave \cite{Boehm:2013jpa}.  

Finally, we comment on how the projected upper-limits compare to a typical annihilation cross
section for a pure $p$--wave annihilating thermal relic in the Galactic center. For this estimate we assume that the DM velocity distribution can be
described by a Maxwell-Boltzmann distribution in the early universe. At
freeze-out ($m_\chi / T \approx 20$) the DM velocity dispersion is
$\sigma_{v,\mathrm{\,fo}} \approx 0.4 c$, where $c$ is the speed of light.
Today, the velocity dispersion in the inner Galaxy is $\sigma_v \approx
4\times10^{-4} c$ \cite{Gultekin:2009qn}.  So $\left<\sigma v\right> =
\left<\sigma v\right>_\mathrm{fo} \left(\sigma_v /
\sigma_{v,\mathrm{\,fo}}\right)^2 \approx
6\times10^{-32}\,\mathrm{\,cm^3\,s^{-1}}$.  Here $\left<\sigma
v\right>_\mathrm{fo}=6\times10^{-26}\mathrm{\,cm^3\,s^{-1}}$ is the $p$-wave
annihilation cross section at the time of freeze out \cite{Diamanti:2013bia}. 
If DM annihilates to monochromatic photons,
this cross-section can be probed 
for $m_\chi \lesssim 10$ MeV.
On the other hand, the projected limits for leptonic final states are at least
one order of magnitude above this reference thermal $p$--wave
cross-section. 

However, we note that in the vicinity of the supermassive
black hole Sagittarius $\mathrm{A^*}$ (\sgr) where much of our signal arises,
the velocity dispersion could potentially be an order of magnitude larger,
increasing the annihilation cross section around the black hole by two orders
of magnitude \citep[see e.g.~][]{Fields:2014pia, Shelton:2015aqa}, potentially
bringing a thermal relic that annihilates through pure $p$--wave processes within reach.

%-----------------------------------------------------------------------------------------
% DISCUSSION AND CONCLUSION
%-----------------------------------------------------------------------------------------
\section{Discussion}
\label{sec:discussion}

In the previous section we have shown that -- under reasonable assumptions for
the systematic uncertainties for the backgrounds/foreground ($\sim
2\%$--$15\%$) and for a reasonable benchmark cosmic-ray transport model --
future MeV \gr missions have the capability to either detect a signature of
thermal relic DM in the case of $p$-wave annihilation (monochromatic photons),
or set stringent bounds that are competitive with current CMB bounds for
$s$-wave annihilation, and much stronger in the case of $p$-wave annihilation.
The existing current indirect detection limits on MeV dark matter can be
improved by two to three orders of magnitude in case of diffuse \grs, and by a
factor of a few in the case of direct measurements of positrons.

Let us now compare the projected bounds derived in this work with
previous studies, comment on the impact of our assumptions and
discuss a potential background from nuclear de-excitations.

\medskip

\emph{Comparison to previous works.} As mentioned in Sec.\ref{sec:results}, our
projections imply that future experiments can improve on existing limits from
Refs.~\cite{Boddy:2015efa,Essig:2013goa} by 2--3 orders of magnitude. 
This difference is partially due to the fact that we consider a better region
of interest (the inner Galaxy), and a more powerful instrument, but mostly
because existing limits were derived without any background subtraction.
In the leptonic annihilation case, another difference between this work and
Ref.~\cite{Essig:2013goa} is that we consider all signal components, both
primary and secondary. This is very important since, in the the inner Galaxy,
secondaries can actually provide the dominant \gr~ signal.  However, it should
be pointed out that Ref.~\cite{Essig:2013goa} considers higher Galactic
latitudes, where in general gas densities are lower and therefore timescales
for gas-related radiative processes longer. As such, a wind or enhanced
diffusion can more easily reduce the secondary signal and enhance the
importance of FSR.

When comparing our projections, based on inner Galaxy observations, to those
from Ref.~\cite{Boddy:2015efa}, 
we find that our results only differ by a factor of a few
from their projected limits for the Draco dwarf-spheroidal galaxy. 
This is merely
coincidental. The ratio of integrated J-factors between Draco and our ROI is
$J_\mathrm{inner\text{--}Galaxy} / J_\mathrm{Draco} \sim 10^{5}$.  However, the
expected background in a dwarf galaxy is only a handful of photons, allowing
for very strong limits despite a limited signal.  Their `diffuse' projections
are however two orders of magnitude weaker.  We believe this difference can be
attributed to significant differences in the ROI as well as different
assumptions on the expected instrumental and background systematics.  More
specifically, Ref.~\cite{Boddy:2015efa} assumes a 15\% systematic uncertainty
on the backgrounds per bin, which is similar to our value for uncertainties
with a large-range correlation in energy space.  However, we adopt only a 2\%
systematic uncertainty for fluctuations that have the width of $\gamma$--ray
lines, as motivated by previous \Fermi-LAT searches~\cite{Albert:2014hwa, Ackermann:2015lka}.

\emph{Assumptions on systematics.} As outlined in Sec.~\ref{sec:sensitivity},
we include systematics in our Fisher forecast and assume a 2\% uncertainty on
small-scales (1\% in energy), and 15\% on large scales (corresponding to 0.5
dex in energy).  These parameters are chosen such that they mimic our
understanding of systematics with \Fermi--LAT today.  We checked that varying
the correlation lengths of the small- and large-scale correlations only has a
small impact on our results.  As mentioned above, the projected sensitivity to
DM annihilation signals in this work are systematics limited. Consequently,
doubling the observation time only yields minor ($\sim10$--$20\%$) improvements
in the limits.

\emph{Cosmic-ray propagation.} Our secondary signal depends on the propagation
model of MeV leptons.  Understanding CR propagation in this low-energy domain
is extremely challenging, and little or no constraints arise from available
data.  In the simplest benchmark case, we model the diffusion coefficient as a
single power law in rigidity and extrapolate down from the GeV domain where
secondary-to-primary ratios set the normalization and slope.  In this case,
diffusion hardly impacts MeV DM (see Fig.~\ref{fig:timescales}), and CR
transport is dominated by the competition between advection and energy losses. 

We also consider a different scenario characterized by a significantly faster
diffusion below 1 GeV. A viable physical picture behind this scenario is e.g.
the dissipation of the magneto-hydrodynamic waves responsible for CR
confinement due to the resonant interaction with low-energy CRs themselves (see
  e.g.  \cite{Ptuskin2006}).  However, even if the diffusion coefficient
  flattens at rigidity of $\sim 1$ GeV due to this effect, our results are not
  severely affected, since cooling and advection timescales are still faster in
  the energy range we are interested in.

In the absence of a wind, the secondary spectra for low DM masses will be
enhanced by a factor of a few, since particles no longer escape their injection
site, but instead cool and radiate in-situ.  The projected limits for this
scenario increase marginally, since secondary emission becomes yet more
important.  For a comparison of the projected limits with and without wind see
Appendix~\ref{sec:cr_transport}.  Note that our benchmark model that includes
an advective wind of $250 \mathrm{\,km\,s^{-1}}$ starting from the plane
appears conservative, since the wind velocity, in particular close to the
plane, could be much smaller \cite{Maurin:2002hw}, thereby increasing the
advection timescale.  However, as argued above, our overall results do not
critically depend on the exact transport parameters chosen.

\emph{$\gamma$--ray lines from nuclear de-excitations.}
A potential background that can produce a sharp feature in the low-energy $\gamma$-ray spectrum and mimic a DM annihilation signal can come from the inelastic interactions of 
low-energy cosmic rays with interstellar gas. This process can produce a rich
spectrum of de-excitation nuclear \gr~ lines from $\sim 0.1$ to $\sim 10$ MeV, spatially correlated with
the gas distribution \cite{Benhabiles-Mezhoud2013}. 
For this reason, a careful morphological study of a possible future signal will be needed in order to disentangle astrophysical and DM interpretations.

\emph{DM profile.} We briefly comment on our choice of dark-matter profile.
This work applies a standard NFW profile with an inner slope of $\gamma=1$.
However, the actual density profile of the Milky--Way is still uncertain as
its determination requires detailed knowledge of the baryonic-mass distribution
\citep[e.g.~][]{Nesti:2013uwa, Pato:2015dua}.
Since annihilation scales as the density squared, having a different profile,
such as the Einasto profile \cite{Einasto:1965czb} or a cored profile, can lead
to relevant differences. For the prompt signal, the limits will simply scale
with the difference in J-factor in our ROI, which for Einasto is $1.5\times$
larger and for a cored-isothermal profile ($r_\mathrm{core}=4.4\mathrm{\,kpc}$)
$10\times$ smaller \cite{Cirelli:2010xx}.  For the secondaries this dependence
is not one-to-one, due to the effects of diffusion, cooling and radiation, of
which the latter two depend on the ambient medium.  However, no large
deviations are expected, in particular since particles are quickly moving away
from the densest gas region in our analysis due to the wind.

\emph{511 keV emission.} Finally, let us briefly comment on 511 keV emission.
When positrons are injected, there will be 511 keV line emission.  However, for
the projected upper limits on the cross section, the associated 511 keV signal
is only a fraction of the total line emission. This should be obvious, since it
was already pointed out by Ref.~\cite{Beacom:2005qv} that in order to explain
the full 511 keV signal with DM there should be an observable in-flight
annihilation signal. Since there are many candidates to produce the low-energy
positrons that produce the 511 keV signal \citep[see][for a
review]{Prantzos:2010wi}, the uncertainty in what fraction could be due to DM
is large. Under the assumption that most 511 keV emission is astrophysical in
origin, it is virtually impossible to detect DM through this channel in the
Inner Galaxy.  On the other hand, we have demonstrated that both prompt and
secondary emission resulting from DM annihilation can produce spectral
signatures that should be detectable in diffuse \grs where the fore- and
backgrounds are better understood. Therefore, if dark matter injects positrons
into the Galactic medium, diffuse \grs offer the best channel for detection.

\section{Conclusions}
\label{sec:conclusion}

We have studied the prospects of indirectly detecting MeV DM by looking for sharp spectral features with a future MeV \gr mission. 

We considered both prompt and secondary emission, for a large range of kinematically accessible two-body final states.
We focused on emission from
the inner Galaxy, where the effect of secondary \grs is especially important.
Using \texttt{DRAGON}, we have consistently modeled for the first time all the relevant energy-loss and radiative processes relevant for leptonic final states at MeV energies.
In addition, we considered the impact of diffusion and advection ($v = 250\mathrm{\,km\,s^{-1}}$), with the latter dominating cosmic--ray transport.
We found that secondary \gr\ emission, in particular in--flight
annihilation and bremsstrahlung, will here often dominate the signal if
$m_\chi \lesssim \mathcal{O}(100\mathrm{\,MeV})$ (see Fig.~\ref{fig:limits_leptons2}).

Projected upper limits for future MeV \gr missions, like the currently proposed
e-ASTROGAM,
were derived using a new approach based on Fisher forecasting with Poisson
likelihoods.  To this end, we assumed that background systematics for the
future instruments will be at a similar level as for \Fermi-LAT today.

For annihilation into \emph{leptonic} final states, we found
(Fig.~\ref{fig:limits_leptons}) 
that diffuse \grs from the inner Galaxy can be used to probe the
annihilation cross-section of sub-GeV DM at a
level that is just a factor of a few weaker than current CMB constraints (for
$s$-wave annihilation and standard halo profiles; for $p$-wave annihilation CMB
constraints vanish and \grs are the stronger probe).  
\emph{Prompt} \gr signals, such as monochromatic photons or box-like spectra, can
be stringently tested throughout the entire MeV--GeV energy range, superseding
current CMB limits by at least an order of magnitude 
(see Fig.~\ref{fig:limits_pions}). Remarkably, values of $\sigma v$ corresponding to thermal $p$--wave production in the early Universe can be probed below DM masses of
$m_\chi\lesssim 20 \mathrm{\,MeV}$.

In summary, future MeV missions such as proposed e-ASTROGAM have in general the
potential to probe large parts of previously unexplored parameter space for
annihilating sub-GeV DM, either improving existing constraints by at least one and up
to several orders of magnitude, or detecting a sharp spectral feature possibly originating from DM annihilation.

%-----------------------------------------------------------------------------------------
% ACKNOWLEDGEMENTS
%-----------------------------------------------------------------------------------------

\acknowledgments
We thank Gianfranco Bertone, Torsten Bringmann, Francesca Calore, Thomas Edwards, Mario Nicola Mazziotta, Christopher McCabe and Nicholas Rodd for useful discussions.
SURFSara is thanked for use of the Lisa Compute Cluster.
This research is funded by NWO through a GRAPPA-PhD fellowship (RB) and through an
NWO VIDI research grant (CW).

%-----------------------------------------------------------------------------------------
% BIBLIOGRAPHY
%-----------------------------------------------------------------------------------------

% \bibliographystyle{apsrev4-1}
\bibliographystyle{JHEP}
\bibliography{papers/lib}

%-----------------------------------------------------------------------------------------
% APPENDIX
%-----------------------------------------------------------------------------------------
\clearpage
%\onecolumngrid
\appendix

%----------------------------
% Annihilation Spectra
%----------------------------

\section{Annihilation spectra}
\label{sec:functions}

Below we present analytic expression for the annihilation spectra of in-flight
annihilation and final-state radiation.

% --- IfA ---%
\subsection{In-flight annihilation}

The differential annihilation cross section for a positron on an electron at
rest is \citep[see][]{1969Ap&SS...3..579S, Beacom:2005qv}
\begin{equation}
  \label{eq:ifa}
  \frac{d\sigma}{dE_\gamma} = \frac{\pi r_e^2}{m_e\gamma^2\beta^2}
  \left(\frac{-\frac{3+\gamma}{1+\gamma} + \frac{3+\gamma}{k} - \frac{1}{k^2}}
  {\left(1 - \frac{k}{1+\gamma}\right)^2}
  - 2
  \right),
\end{equation}
for $\gamma(1-\beta)\leq 2k - 1\leq \gamma(1+\beta)$.  Here $r_e$ is the
classical radius of the electron, $m_e$ is the electron mass $\gamma$ is the
Lorentz factor, $\beta = v/c$ and $k=E_\gamma / m_e$.  From the kinematic
bounds we can see that For non-relativistic positrons, i.e.~$\gamma\rightarrow
1$ and $\beta\rightarrow 0$, photons from IfA are mono-energetic at $E_\gamma =
m_e$. In the ultra-relativistic limit ($\gamma\rightarrow\infty$ and
$\beta\rightarrow 1$) we get that $\gamma\left(1 - \beta\right.)\rightarrow 0$.
As such there is a lower limit to the photon energy of $m_e/2\leq E_\gamma $.

The production rate of photons from in-flight annihilation by a single positron
is then,
\begin{equation}
  \frac{dN}{dE_\gamma\,dt} = N_{e^-}\beta c\frac{d\sigma}{dE_\gamma}.
\end{equation}
Here $N_{e^-}$ is the target density of electrons. Note that the differential
cross section (Eq.~\ref{eq:ifa}) is already weighted with a photon multiplicity
of 2.

\subsection{Final-state radiation}

We present the expressions for final-state radiation applied in this paper.

\paragraph{Direct annihilation into $e^+e^-$.} In case of direct annihilation
into fermions, the number of FSR photons per annihilation is:
\cite{Beacom:2004pe,Bringmann:2007nk,Essig:2009jx}:
\begin{equation}
\begin{split}
\label{eq:fsr1}
\frac{dN}{dE_\gamma} 	&= \frac{1}{\sigma_\mathrm{tot}}\frac{d\sigma_{FSR}}{dE_\gamma} \\
				&= \frac{\alpha}{\pi} \frac{1}{E_\gamma} \left[\ln\left(\frac{s'}{m_e^2}\right)-1\right]
                	\left[1 + \left(\frac{s'}{s}\right)^2\right],
\end{split}
\end{equation}
with $\sigma_\mathrm{tot}$ the total annihilation cross section into $e^+e^-$,
$\alpha=1 / 137$ the fine-structure constant, $s=4m_\chi$ (roughly the
center-of-mass energy for non-relativistic DM) and $s' = 4m_\chi(m_\chi -
E_\gamma)$.

\paragraph{Cascade annihilation} For a one-step cascade annihilation,
$\chi\chi\rightarrow\phi\phi$, where subsequently $\phi\rightarrow e^+e^-$, the
FSR spectrum is a little more complicated since the mediators, $\phi$, are
boosted. An analytic expression can be derived by starting from
Eq.~\ref{eq:fsr1}, but now for $\phi\rightarrow e^+e^-$, so $s=m_\phi^2$ and
then boosting back to the rest frame of the DM.  The resulting expression is
\cite{Essig:2009jx},
\begin{equation}
\begin{split}
\frac{dN}{dE_\gamma} = \frac{2\alpha}{\pi E_\gamma}
	& \left\{
	x^2 + 2 x \left[\mathrm{Li_2}\left(\frac{m_\phi - 2 m_e}{m_\phi - m_e}\right) - \mathrm{Li_2}(x)\right]
    + (2 - x^2)\ln(1 - x) + \right.\\
    &
    \left[\ln\left(\frac{m_\phi^2}{m_e^2}\right)-1\right]
    \left[2 - x^2 +2x\ln\left(\frac{\left(m_\phi - m_e\right)x}{m_\phi -2m_e}\right)
    - \frac{\left(m_\phi^2 - 2 m_e^2\right)x}{\left(m_\phi - m_e\right)\left(m_\phi - 2 m_e\right)}\right] -\\
    &
    \frac{x}{2m_e^2 -3m_\phi m_e + m_\phi^2} \left[
    2m_e^2\left(2 - \ln\left(\frac{m_e^2 x^2}{(m_\phi -2 m_e)^2(1-x)}\right)\right) \right. \\
    & - \left. \left.
    3 m_e m_\phi \left(\frac 4 3 - \ln \left(\frac{m_e (m_\phi -m_e)x^2}{(m_\phi - 2m_e)^2(1-x)}\right)\right) \right.\right. \\
    & + \left. \left. m_\phi^2\left(1 - \ln\left(\frac{(m_\phi - m_e)^2x^2}{(m_\phi - 2 m_e)^2(1 -x)}\right)\right)
    \right]
    \right\}
\end{split}
\end{equation}
where $x = E_\gamma / m_\chi$ and 
$\mathrm{Li_2}(z)=\int_z^0 \frac{\ln(1-t)}{t} dt$ is the dilogarithm.

%----------------------------
% Cosmic ray transport
%----------------------------

\section{Cosmic-ray transport}
\label{sec:cr_transport}

\begin{figure*}[h]
    \centering
    \begin{subfigure}{0.45\textwidth}
        \includegraphics[width=\textwidth]{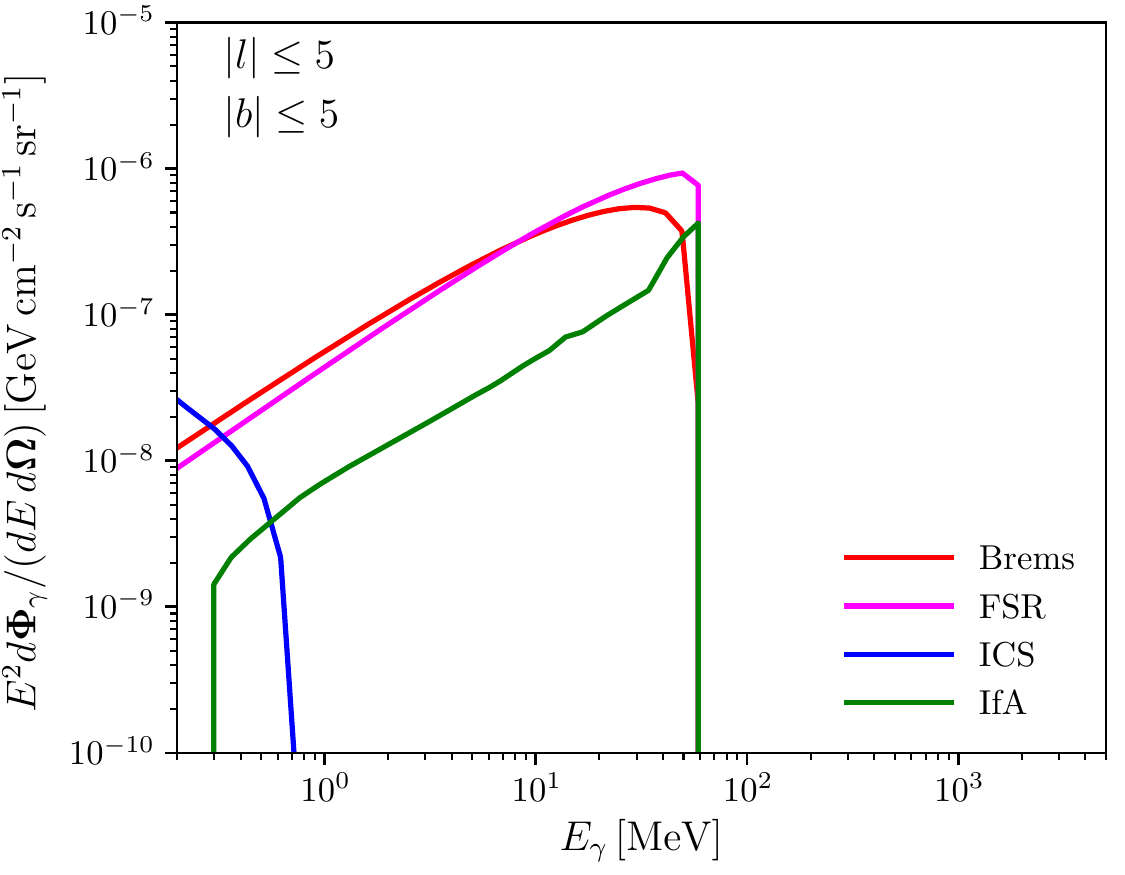}
        \caption{Benchmark model: slow diffusion with wind.}
        \label{fig:speca}
    \end{subfigure}
    ~ 
    \begin{subfigure}{0.45\textwidth}
        \includegraphics[width=\textwidth]{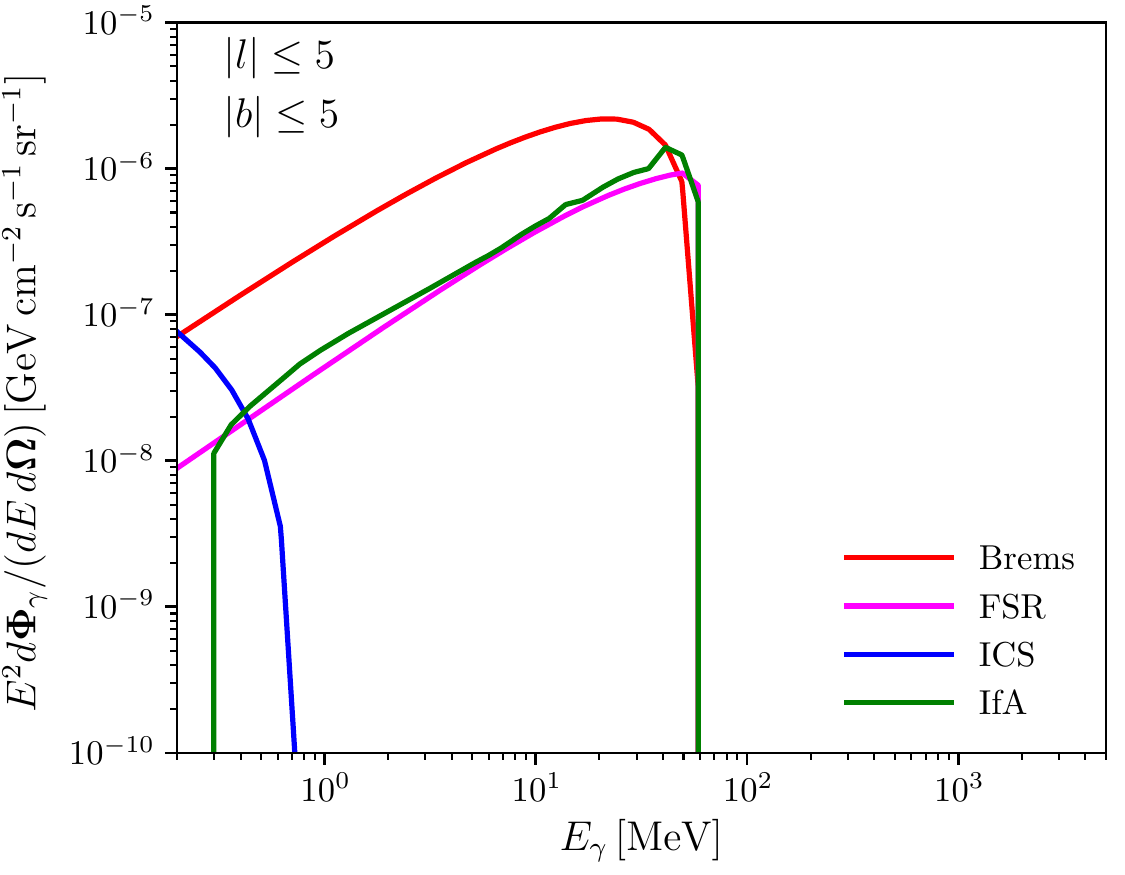}
        \caption{Slow diffusion, no wind.}
        \label{fig:specb}
\end{subfigure}
    ~
    \begin{subfigure}{0.45\textwidth}
        \includegraphics[width=\textwidth]{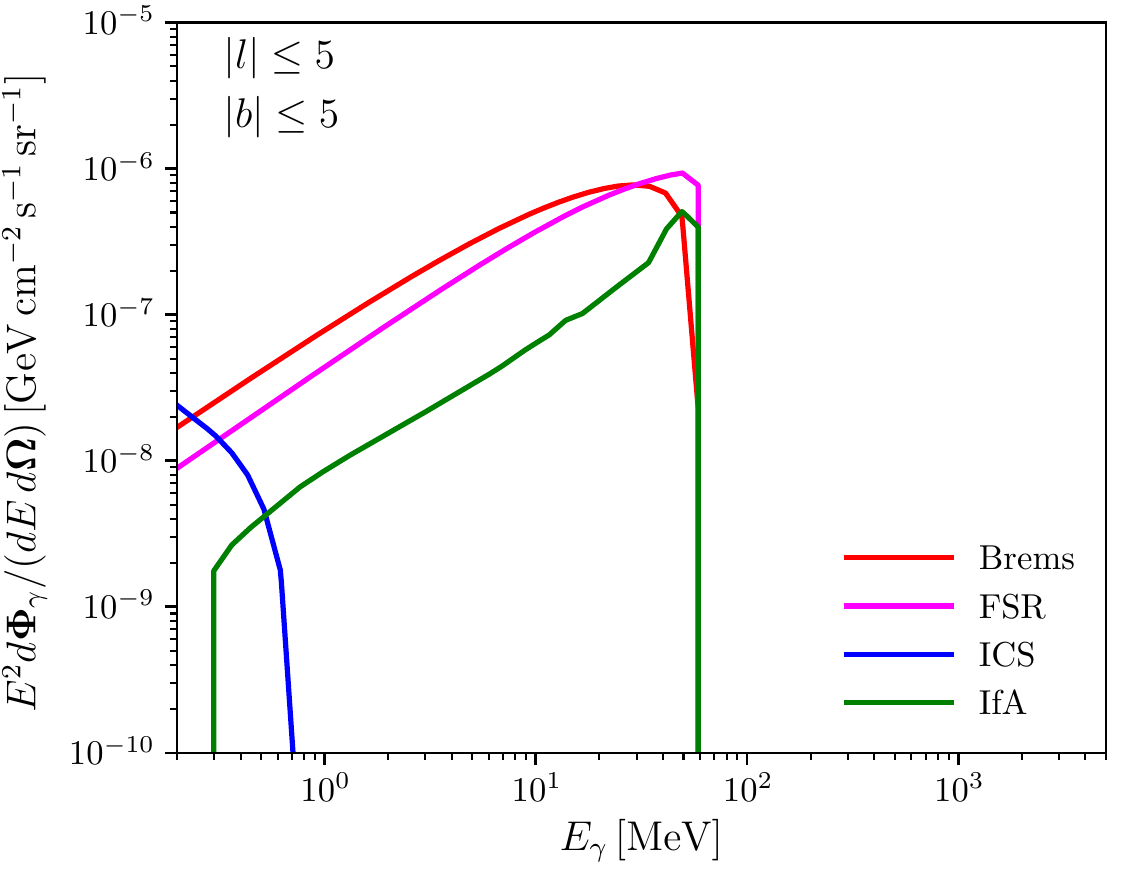}
        \caption{Fast diffusion and wind.}
        \label{fig:specc}
\end{subfigure}
    ~
    \begin{subfigure}{0.45\textwidth}
        \includegraphics[width=\textwidth]{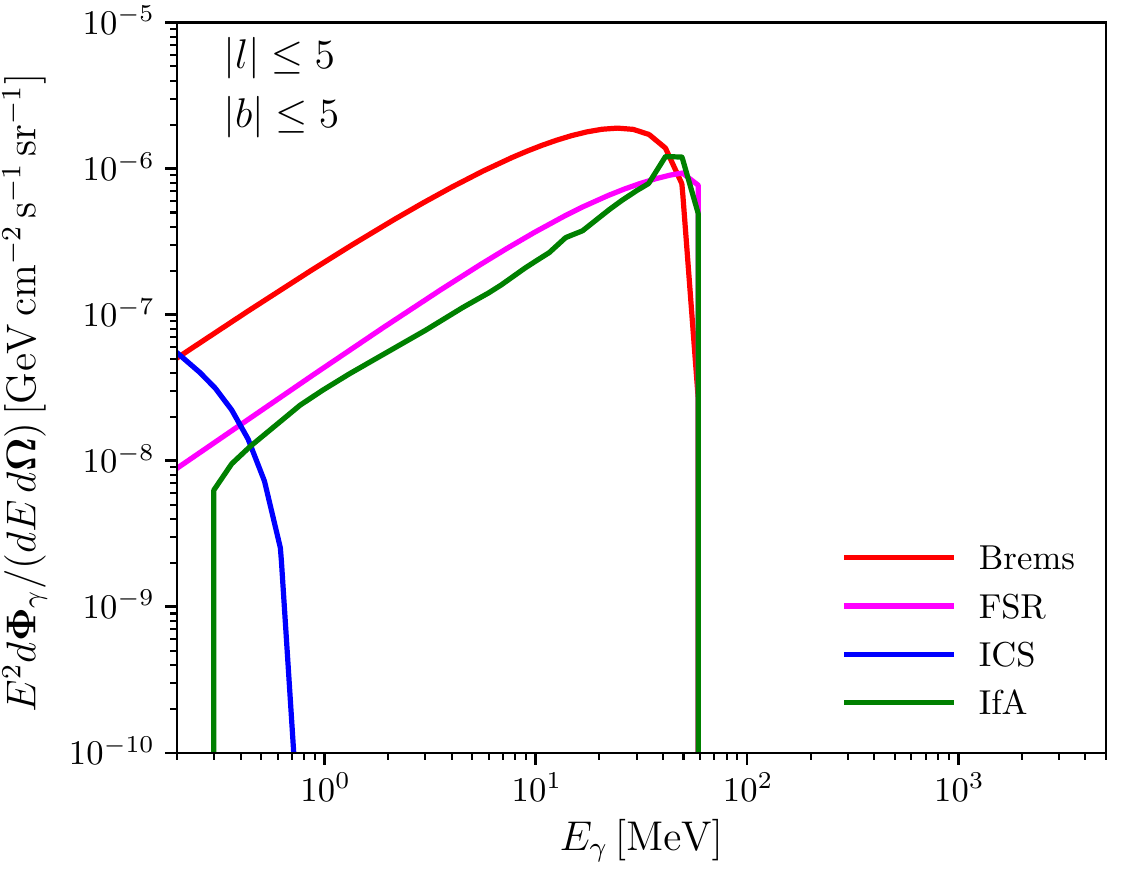}
        \caption{Fast diffusion, no wind.}
        \label{fig:specd}
    \end{subfigure}
    \caption{\label{fig:spectra_transport}
    Spectra for $m_\chi=60\mathrm{\,MeV}$, 
    $\chi\chi\rightarrow e^+e^-$ and 
    $\left<\sigma v\right> = 10^{-28}\mathrm{\,cm^3\,s^{-1}}$ 
    and considering different cosmic-ray transport scenarios.}
\end{figure*}

\begin{figure*}[h]
    \centering
    \begin{subfigure}{0.45\textwidth}
        \includegraphics[width=\textwidth]{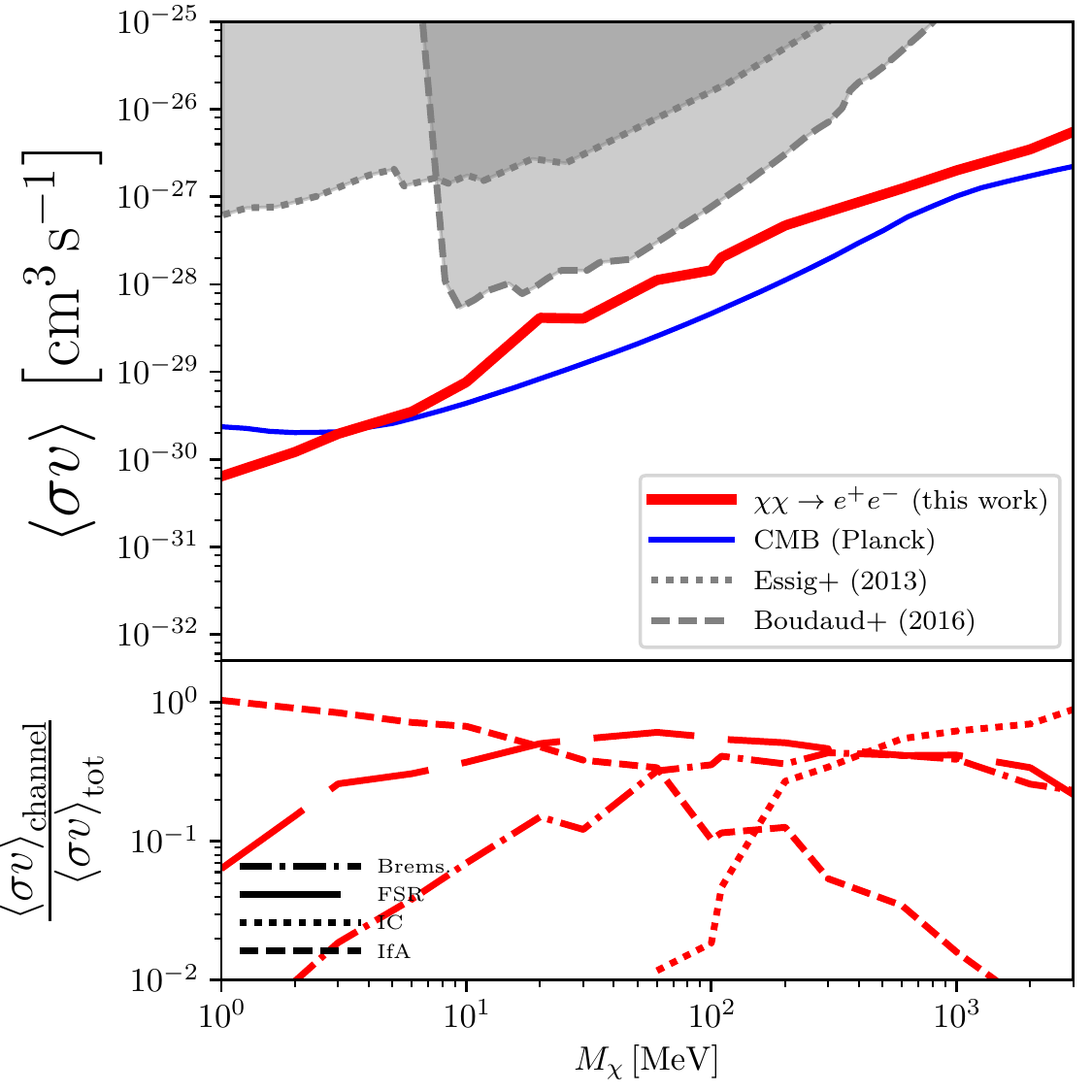}
        \caption{Benchmark model: slow diffusion with wind.}
    \end{subfigure}
    ~ 
    \begin{subfigure}{0.45\textwidth}
        \includegraphics[width=\textwidth]{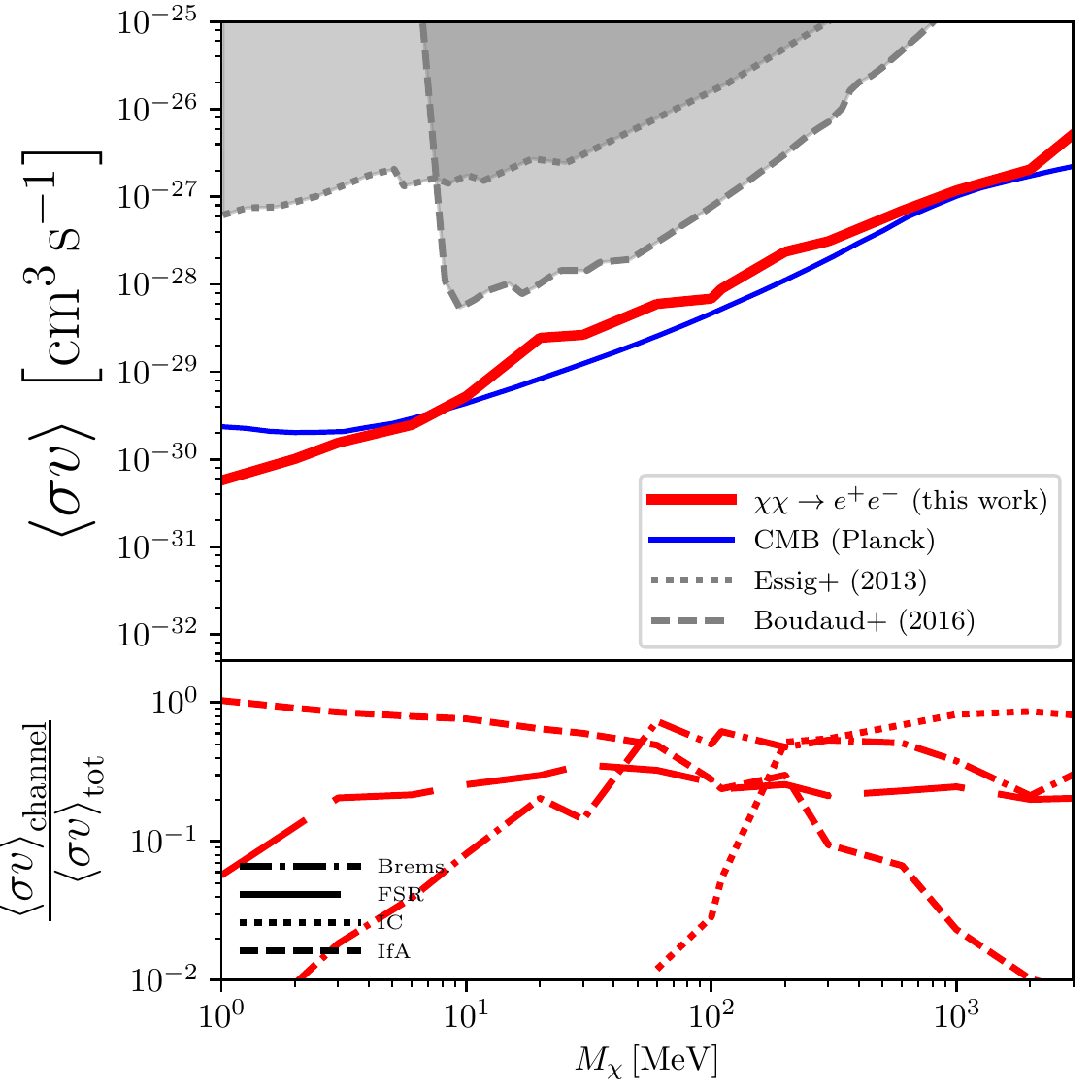}
        \caption{Slow diffusion, no wind.}
      \end{subfigure}
    \caption{\label{fig:limits_transport} Projected upper limits for
  $\chi\chi\rightarrow e^+e^-$ for different cosmic-ray transport scenarios.
Figures identical to Fig.~\ref{fig:limits_leptons}, but excluding the cascade
and muon channel. }
\end{figure*}

In this appendix we discuss the results for different setups for the cosmic-ray
transport.  Corresponding limits for $\chi\chi\rightarrow e^+e^-$ are shown in
Fig.~\ref{fig:limits_transport}.  In addition, we also show how the secondary
spectra change when the transport parameters are varied in
Fig.~\ref{fig:spectra_transport}.  Our benchmark scenario includes an advective
wind perpendicular to the plane with a velocity of
$v_\mathrm{wind}=250\mathrm{\,km\,s^{-1}}$ and confined within the inner 3 kpc
radius. The diffusion coefficient was modeled as a single power law.  However,
the behavior of the diffusion coefficient at low energies is unconstrained.  We
also study the case where there the diffusion coefficient flattens below 1.5
GeV, leading to faster diffusion at low energies.  In this scenario, the
diffusion coefficient as a function of momentum is modeled as a broken power
law,
\begin{equation}
  D(p) = \left\{
  	\begin{array}{lr}
    D_0 \left(\frac{p_b}{\rho_0}\right)^{\delta}
    \left(\frac{p}{p_b}\right)^{\delta_l} &    
    \text{ for } p< p_b \\
    D_0 \left(\frac{\rho}{\rho_0}\right)^\delta &
    \text{ for } p_b \leq p,
    \end{array}
    \right.
\end{equation}
where everything is as in Eq.~\ref{eq:diffusion}, with the addition of the
break momentum, $p_b=1.5 \mathrm{\,GeV}$, and the the diffusion coefficient
becomes energy independent below the break, i.e.~$\delta_l = 0$ (see
Fig.~\ref{fig:timescales}).  In addition, we consider scenarios in which there
is no wind.

Figure \ref{fig:spectra_transport} shows the spectra for a $m_\chi=60
\mathrm{\,MeV}$ DM particle annihilating to an $e^+/e^-$ pair. Figure
\ref{fig:speca} is identical to Fig.~\ref{fig:spectrum} and depicts our
benchmark model negligible diffusion at low momenta and an advective wind. In
Fig.~\ref{fig:specb} the wind has been turned off.  Naturally,
final-state-radiation, being prompt, emission is unaffected.  The secondary
emission increases by a factor of a few compared to Fig.~\ref{fig:speca}.
Without advection and with single-power law diffusion the electrons and
positrons loose their energy almost in-situ. Whereas in the presence of a wind
they are likely to be advected out of our ROI before radiating.  In the bottom
two panels we show the results for the scenario in which the diffusion
coefficient flattens below momenta of 1.5 GeV, again with
(Fig.~\ref{fig:specc}) and without (Fig.~\ref{fig:specd}) wind. As can be seen,
this has little impact on the result since in either the wind dominates
transport as is the case in the bottom left plot, or cooling timescales are
still faster than diffusion timescales (both bottom panels). All this is
evident from Fig.~\ref{fig:timescales}

In addition, we compare projected upper limits the scenarios with and without
wind and for single power-law diffusion in Fig.~\ref{fig:limits_transport}.  We
do not show limits for the scenario with a flat diffusion coefficient below 1.5
GeV as this only marginally affects the results, as discussed above.  All
assumptions on the instrument and background are identical to those in the main
text.  As mentioned above, in absence of the wind the secondary particles cool
and radiate almost in-situ. Resulting in secondary emission being yet more
important, with IfA being the dominant contribution to the limits upto $m_\chi
\sim 100$ MeV. 

%----------------------------
% Spectral plot
%----------------------------

\section{Spectra for different channels}
\label{sec:spectral_plots}

Below we present example spectra for the different annihilation channels
presented in this paper. All spectra are for benchmark setup and correspond to
a window $|l|, |b| < 5^\circ$.  The annihilation cross section is everywhere
set to $\left<\sigma v\right> = 10^{-28}\mathrm{\,cm^3\,s^{-1}}$. 

\begin{figure*}[h]
	\label{fig:spectra_channels}
    \centering
    \begin{subfigure}{0.45\textwidth}
        \includegraphics[width=\textwidth]{figs/UnconvolvedSpectrum_DM_slow_diffusion_Delta_Mdm60_vwind250_l5_b5_NoBackground.pdf}
        \caption{Spectrum for $\chi\chi\rightarrow e^+e^-$
        and $m_\chi=60\mathrm{\,MeV}.$}
        \label{fig:spec_ee}
    \end{subfigure}
    ~ 
    \begin{subfigure}{0.45\textwidth}
        \includegraphics[width=\textwidth]{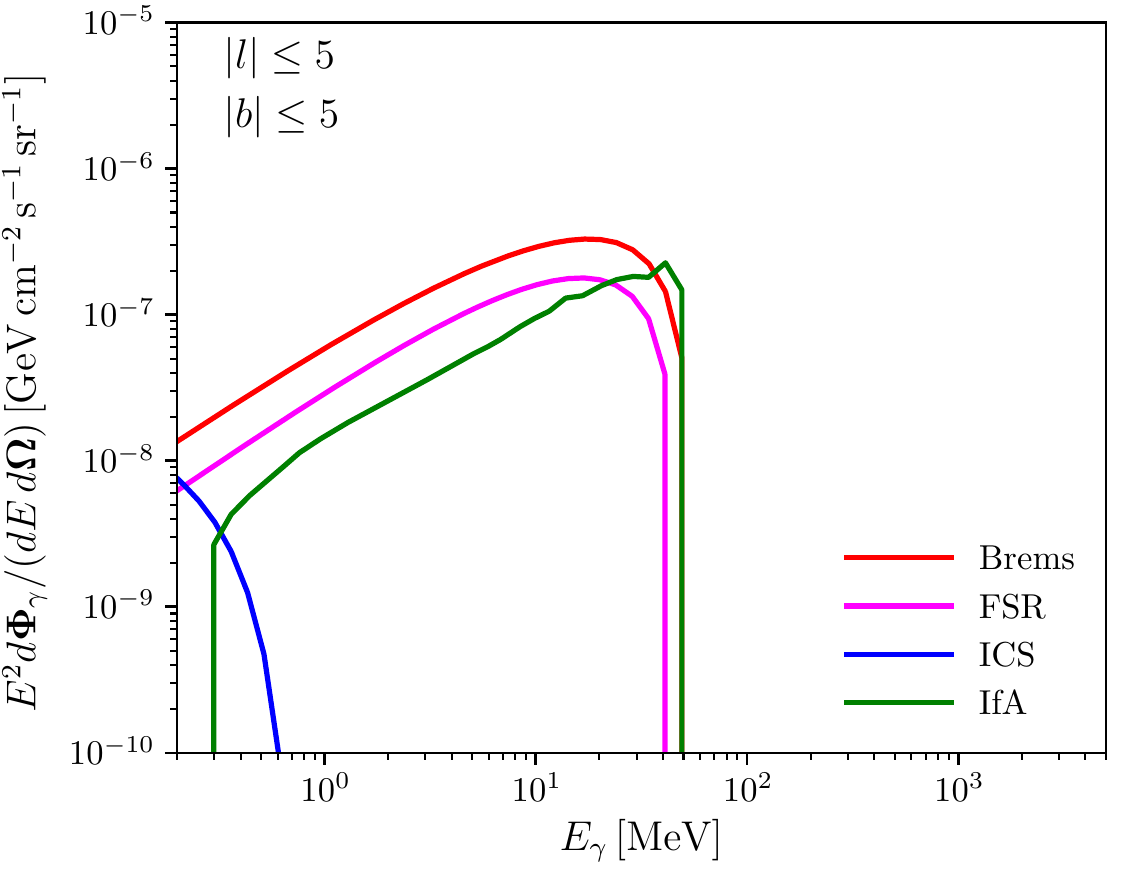}
        \caption{Spectrum for $\chi\chi\rightarrow\phi\phi\rightarrow e^+e^-e^+e^-$,
        $m_\chi=60\mathrm{\,MeV}$ and $m_\phi=5\mathrm{\,MeV}$.}
        \label{fig:spec_cascade}
\end{subfigure}
    ~
    \begin{subfigure}{0.45\textwidth}
        \includegraphics[width=\textwidth]{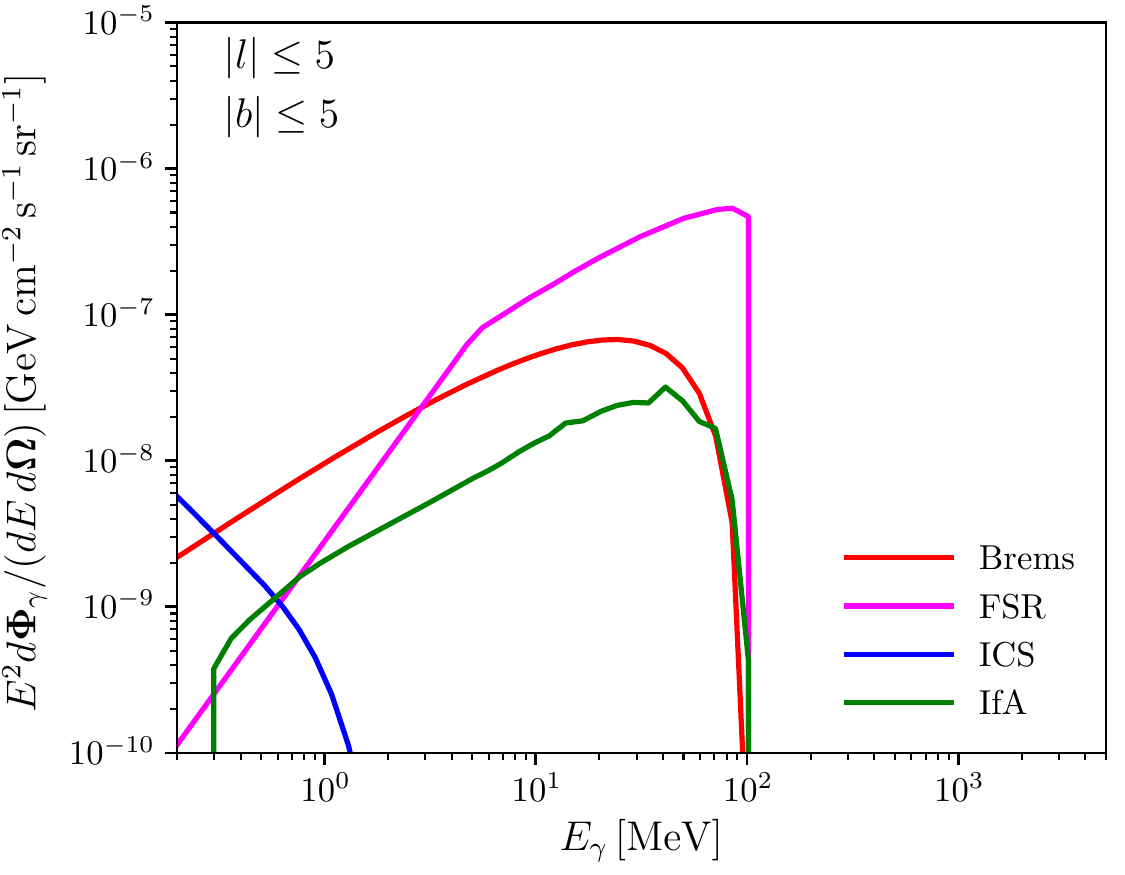}
        \caption{Spectrum for $\chi\chi\rightarrow\mu^+\mu^-$
        and $m_\chi=110\mathrm{\,MeV}.$}
        \label{fig:spec_mm}
\end{subfigure}
    ~
    \begin{subfigure}{0.45\textwidth}
        \includegraphics[width=\textwidth]{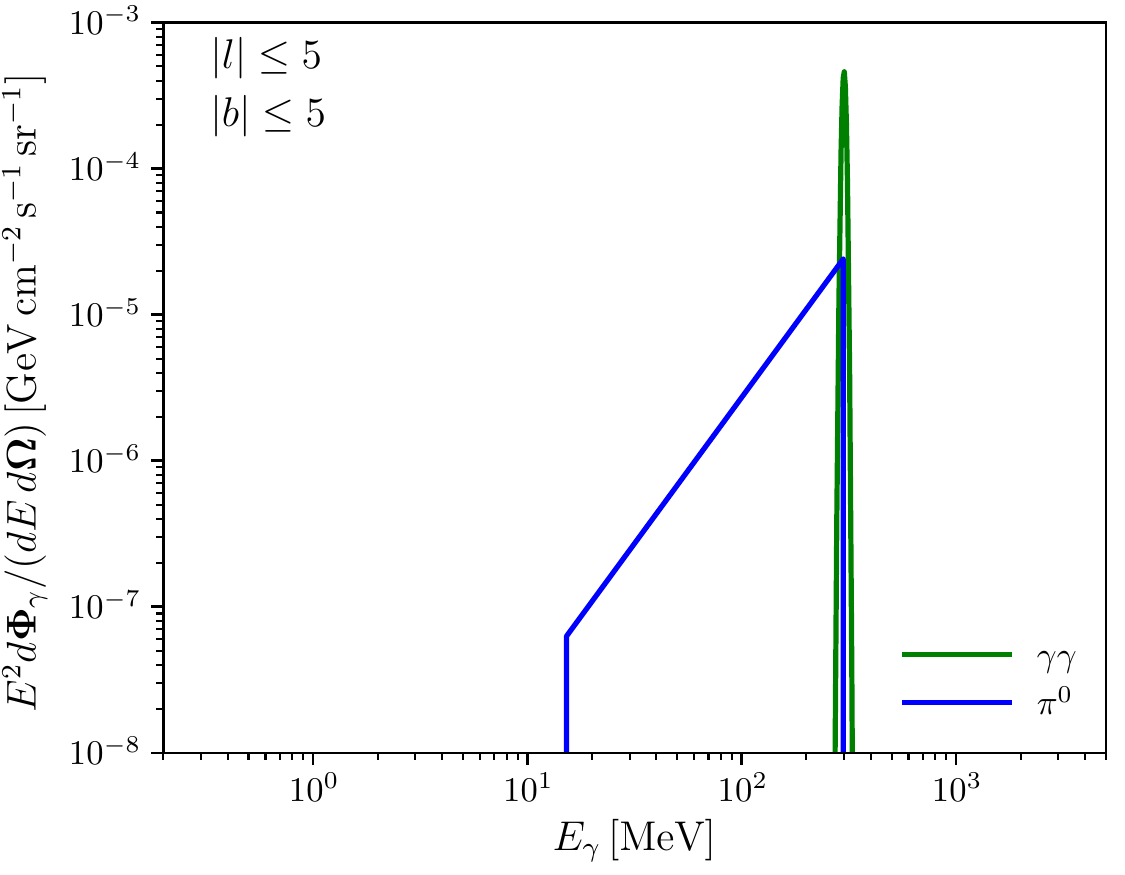}
        \caption{Spectrum for $\chi\chi\rightarrow\pi^0\gamma$
        and $m_\chi=300\mathrm{\,MeV}$. 
        The width of the line is set to 2\% to ease the eye.}
        \label{fig:spec_pi0gam}
    \end{subfigure}
    \caption{Spectra (prompt and secondary) for the channels considered in this
      paper.  Where secondary emission is present, we assumed our benchmark
      cosmic-ray transport model, with slow diffusion and an advective wind of
      $v=250\mathrm{\,km\,s^{-1}}$.  None of these spectra has been convolved
    with the instrumental resolution.  }
\end{figure*}

In Fig.~\ref{fig:spec_ee} we show the spectrum from $\chi\chi\rightarrow
e^+e^-$ for $m_\chi=60\mathrm{\,MeV}$. For the same dark matter mass we show
the spectrum for the cascade channel in Fig.~\ref{fig:spec_cascade}. The
mediator mass is $m_\phi=5\mathrm{\,MeV}$.  The FSR spectrum is now suppressed
compared to the direct annihilation channel due to the smaller mass splitting
between the mediator and and the electron, relative to the mass splitting
between the electron and the dark matter particle.  Also, the secondary spectra
(IfA and bremsstrahlung) become more smooth since the injection spectrum of
electrons and positrons gets spread out as a result of the boost of the
mediator.

In the bottom panels we show the result for the muonic channel
(Fig.~\ref{fig:spec_mm}) at $m_\chi = 110\mathrm{\,MeV}$.  For the muonic
channel FSR is the dominant component, which in this case is generated at two
stages, when dark matter annihilates to a muon pair, and when the muon
subsequently decays. Also, secondaries are more suppressed here compared to the
direct or cascade annihilation channel, since the spectrum gets broadened as a
result of the non-zero momentum of the muon and only one $e^+e^-$ pair is
produced, compared to two in the cascade channel.

Finally, we show the spectrum for $\chi\chi\rightarrow\pi_0\gamma$ at $m_\chi =
300\mathrm{\,MeV}$ in Fig.~\ref{fig:spec_pi0gam}.  The spectrum consists of a
monochromatic line (which we gave a width of 2\% for visual simplicity) and a
box, since the pion is a scalar. The cross section is the same in all figures,
but note the different vertical scale in Fig.~\ref{fig:spec_pi0gam} compared to
the other figures.

\end{document}